\documentclass[
prb,twocolumn,10pt,
superscriptaddress,
 amssymb,amsmath,
 aps,
 footinbib, amsart
]{revtex4-1}

\usepackage{pagenote}
\renewcommand*{\pagenotesubhead}[1]{}

\makepagenote

\usepackage{tcolorbox}
\usepackage{graphicx}
\usepackage{dcolumn}
\usepackage{cancel} 
\usepackage{amssymb}
\usepackage[normalem]{ulem}
\usepackage{bm}
\usepackage{subfigure}

\usepackage{mathtools}
\DeclarePairedDelimiter\evaluat{.}{\rvert}

\begin{document}

\newcommand{\todo}[1]{{\color{cyan}{#1}}}
\newcommand{\ch}[1]{{\color{black}{#1}}}
\newcommand{\AG}[1]{{\color{purple}{#1}}}
\newcommand{\LT}[1]{{\color{green}{#1}}}
\title{Electric field direction dependence of the electrocaloric effect in BaTiO$_3$}

\author{Lan-Tien Hsu}
\email{lan-tien.hsu@ruhr-uni-bochum.de}
\affiliation{%
 Interdisciplinary Centre for Advanced Materials Simulation (ICAMS) and Center for Interface-Dominated High Performance Materials (ZGH), Ruhr-University Bochum, Universitätsstr. 150, 44801 Bochum, Germany
}%
\author{Frank Wendler}
\email{frank.wendler@fau.de}
\affiliation{
Institute of Materials Simulation, Department of Materials Science, Friedrich-\-Alexander University of Erlangen-N\"urnberg, Dr.-Mack-Strasse 77, 90762 F\"urth, Germany
}
\author{Anna Grünebohm$^{1,} $}%
\email{anna.gruenebohm@ruhr-uni-bochum.de}
\altaffiliation[Also at ]{ICAMS and ZGH, Ruhr-University Bochum.}

\begin{abstract} 

Single crystalline ferroelectric perovskites show a large electrocaloric effect at electric field-induced phase transitions, promising for solid-state cooling technologies. However, paraelectric-ferroelectric transition temperatures are often too high for practical applications and lower transitions are so far underrepresented in literature. Particularly, the role of thermal hysteresis and electric field direction on the \ch{caloric response} is critical, especially for polycrystalline materials, but not yet fully understood. 
Using ab initio based coarse-grained molecular dynamics simulations, we show how transition temperatures depend on the direction of the applied field.
Furthermore, we reveal that the choice of electric field direction can reduce thermal hysteresis and can adjust the \ch{temperature ranges where large and reversible caloric responses occur}. Furthermore, we propose a phenomenological descriptor for the qualitative changes in transition temperature with field direction.
\ch{This descriptor is valid for both BaTiO$_3$ and PbTiO$_3$, even though both materials show different microscopic electric field coupling.} Finally, we identify favorable temperature- and texturing conditions for large and reversible caloric responses in polycrystals.

\textbf{keyword: electrocaloric effect, solid-state cooling, ferroelectrics, structural phase transition, molecular dynamics simulations, electric field direction dependence, symmetry, perovskite oxides, thermal hysteresis} 
\end{abstract}

\maketitle

\section{Introduction}
\ch{The electrocaloric effect (ECE) is promising for solid-state cooling technologies.\cite{ozbolt_electrocaloric_2014, aprea_comparison_2017, shi_electrocaloric_2019, wang_high-performance_2020, greco_electrocaloric_2020, Torello.2022} It is the electric field-induced adiabatic temperature change ($\Delta T_{\text{ad.}}$)  by the 
 reallocation of entropy between lattice vibrations and the dipole configuration. 
This temperature change depends on the change in polarization ($\vec P$) with temperature ($T$) and with the change of the applied external electric field (from $\vec{E}_\text{in}$ to ${\vec{E}}_\text{fi}$). 
For the sake of simplicity, we use the term ``field" for the electric field and ``polarization" for the vector field of polarization.
For continuous, ergodic polarization changes, the adiabatic temperature change is given by the thermodynamics Maxwell relation as
 \begin{equation}
\begin{split}\label{eq:TD_P}
\Delta T_\text{ad.} &= - \int_{{\vec{E}}_\text{in}}^{{\vec{E}}_\text{fi}} \frac{T}{C_{{{E}}}(T)} \cdot \evaluat* {\frac{\partial \vec{P}}{\partial T} }_{{\vec{E}}} \;d{{\vec{E}}}\;,\\
\end{split}
\end{equation}
with the heat capacity $C_{{E}}(T)$ and the pyroelectric coefficient $\evaluat* {\frac{\partial \vec{P}}{\partial T} }_{{\vec{E}}}$.
\begin{figure}[h!]
    \centering
    \includegraphics[width=0.45\textwidth]{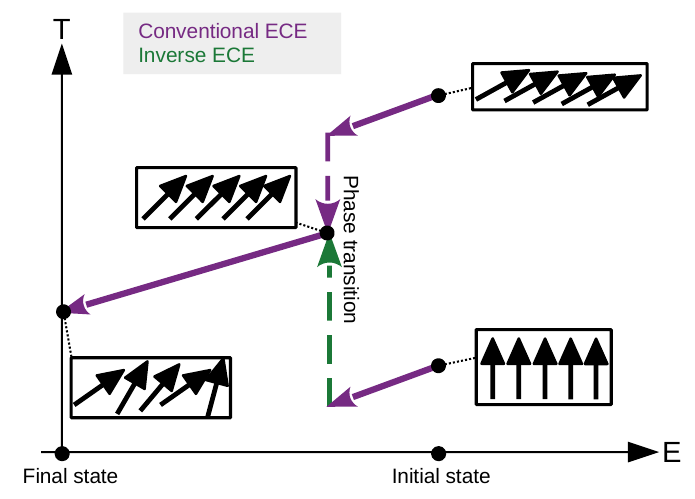}
    \caption{\ch{Sketch of conventional (purple) and inverse (green) adiabatic temperature ($T$) changes, i.e.\ ECE, under electric field ($\vec{E}$) removal. The dipoles are represented by black arrows. In the initial state, the systems are pre-poled by an external electric field which induces a high degree of ordering of the dipoles. This ordering is reduced under field removal, inducing a conventional ECE, i.e.,\ the material cools down under adiabatic conditions, as illustrated by solid purple arrows. If the system crosses a first-order phase transition, its latent heat results in an additional temperature change, see Eq.~\eqref{eq:TD_CC}, as illustrated by dashed arrows. This response may be conventional or inverse if the system enters the high- or low-temperature phase during field removal, respectively. 
    }
   }
    \label{fig:overviewsketchECE}
\end{figure}
At first-order phase transitions, the adiabatic temperature change contributed by latent heat is given by the Clausius--Clapeyron equation as
  \begin{equation}
\begin{split}\label{eq:TD_CC}
\Delta T_\text{ad.}^\text{LH} &= 
- \frac{T}{C_{{E}}(T)} \Delta \vec{P}({\vec{E}}_\text{t}(T))\cdot \left| \frac{d{\vec{E}}_\text{t}}{dT} \right|,
\end{split}
\end{equation} 
where ${\vec{E}}_\text{t}$ is the field strength to induce the transition at the given temperature. \cite{Marathe.2017,Moya.2014, Mischenko.2006, Neese.2008,Grunebohm.2018}

The ECE is called conventional when the temperature decreases under electric field removal. This conventional ECE can be induced within one phase by a field-induced change of dipole ordering, see solid purple arrows in Fig.~\ref{fig:overviewsketchECE}. A conventional ECE involving latent heat occurs if the field stabilizes a phase with lower entropy against one with higher entropy.  For example, a conventional ECE can be observed when the material transits from the orthorhombic (O) to the tetragonal (T) phase during field removal, which leads to an abrupt decrease in material's temperature, see dashed purple arrow in Fig.~\ref{fig:overviewsketchECE}.
 On the other hand, the ECE is called inverse when adiabatic cooling can be achieved under field application, while the material heats up under field removal.\cite{Wu.2017, Grunebohm.2018} This is possible at phase transitions if the field direction stabilizes the phase with higher entropy.
 For example, a field along $[001]$ stabilizes the T phase and during field removal the temperature increases abruptly at the T to O transition, as illustrated by the green dashed arrow in Fig.~\ref{fig:overviewsketchECE}.  For further details about the origins of the inverse ECE, we refer the reader to Ref.~\onlinecite{Grunebohm.2018}.

Typical cooling applications require $\Delta T_{\text{ad.}}$ of about 20~K.\cite{meng_electrocaloric_2021}
For thin films, ECE of up to 40~K have been reported experimentally,\cite{lu_organic_2010} which is in the same order of magnitude as in thermoelectric 
\cite{qin_solid-state_2022} and other caloric, e.g.,\ elasto- and magneto-caloric, materials.\cite{hou_materials_2022} 
However, the amount of heat that thin films can transfer is limited. 
Furthermore, these so-called giant values of $\Delta T_{\text{ad.}}$ are associated with electric field variations at the order of MV/cm. Due to lower breakdown strengths, bulk-like materials cannot bear such strong fields\cite{moya_giant_2013} and show maximal experimental values of $\Delta T_{\text{ad.}}$ 2--5.5~K experimentally for field strengths of 10--300~kV/cm.\cite{hou_materials_2022,nouchokgwe_giant_2021,nair_large_2019} 
These temperature changes would be sufficient for cooling devices if regenerator or cascade designs are used.\cite{meng_electrocaloric_2021,meng_cascade_2020} 
But, for these field strengths, the significant caloric effect is typically confined to narrow temperature ranges close to the phase transition with sufficient $\evaluat* {\frac{\partial \vec{P}}{\partial T} }_{{\vec{E}}}$, cf.~Eq.~\eqref{eq:TD_P}.\cite{moya_giant_2013,rose_giant_2012,hou_materials_2022}
This temperature window is only 5~K around 400~K for BaTiO$_3$ under a field of 10~kV/cm,\cite{bai_both_2013} while a temperature window of about 60~K has been reported near 330~K for PbSc$_{0.5}$Ta$_{0.5}$O$_{3}$  multilayer capacitors under a field of 300~kV/cm.\cite{nair_large_2019} For results for other materials, we refer the reader to literature.~\cite{moya_giant_2013,rose_giant_2012, shi_electrocaloric_2019, greco_electrocaloric_2020,mikhaleva_caloric_2012}

To expand the temperature range with a large response in moderate field strengths and to bring it to and below ambient temperatures, remain challenging. One common solution is to reduce the paraelectric-ferroelectric (PE-FE) transition temperature by substitution of the A-site or/and B-site cations\cite{lv_manipulation_2020,uddin_effect_2021,bratton_phase_1967,lisenkov_tuning_2018,grunebohm_optimizing_2018}. 
In addition, the lower temperature ferroelectric-ferroelectric (FE-FE) transitions have gained attention.\cite{Marathe.2017,taxil_modeling_2022} There, sizeable responses related to the latent heat of these first-order transitions have been predicted.\cite{Marathe.2017,taxil_modeling_2022} 
However, first-order phase transitions are also subject to thermal hysteresis. If the system does not fully cross the phase transition during the electric field variation or if the system ends up in a coexistence range, the ECE may be reduced, thermal-history-dependent, and irreversible.\cite{li_effects_2022, Marathe.2018, Moya.2014} Therefore, the smaller the thermal hysteresis, the larger the temperature window with a large and reversible response.

In addition to the adiabatic temperature change, also the coefficient of performance, i.e.\ the ratio of cooling power and power input in ECE devices, has to compete with compressor-based refrigerators. As thermal hysteresis results in energy losses, it has been reported that the ratio between thermal hysteresis and $\Delta T_{\text{ad.}}$ needs to be smaller than 10\% for efficient cooling devices.\cite{hess_thermal_2020} 
Yet, a ratio of $\Delta T_{\text{ad.}}$ and thermal hysteresis without an external field of 500\% has been found in BaTiO$_3$ single crystals\cite{moya_giant_2013} at the PE-FE transition, and thermal hysteresis is even larger at FE-FE transitions.
While reducing thermal hysteresis is the focus of research for other ferroic cooling technologies,\cite{Gutfleisch.2016} it is so far underrepresented in the case of the ECE. 
Concepts to bypass thermal hysteresis include weakening the first-order character of transitions by large electric fields,\cite{Novak.2013, Marathe.2017, Liu.2016b,li_effects_2022} reducing the energy barrier for nucleation and growth of the new phase by domain and defect engineering, and combining different fields, e.g.\ adding mechanical stimuli.\cite{Grunebohm.2021,Gutfleisch.2016} 

At first-order FE-FE transitions, furthermore, functional and structural fatigue caused by the incompatible strain during the lattice deformation challenges the application of ferroelectric single crystals in cooling devices.\cite{bai_both_2013} Improved stability and reliability of the caloric response is possible in polycrystalline ceramics where grain boundaries buffer the stress at the transitions.\cite{bai_both_2013, pramanick_strain_2012,fang_fatigue_2005} Yet, it has been reported by an experimental group that $\Delta T_{\text{ad.}}$ in BaTiO$_3$ ceramics is 32\% smaller than in single crystals.\cite{li_near-room-temperature_2021} 
The fundamental understanding of the caloric responses in polycrystals is so far incomplete and these are underrepresented in theoretical studies.

Present challenges in material design thus include (1) tuning the temperature ranges where sufficiently large $\Delta T_{\text{ad.}}$ occurs to room temperature and broadening these ranges for moderate strengths of the applied field, (2) maintaining a large
response when going from single crystals to technically relevant ceramics, and (3) realizing reversible responses, i.e.,\ reducing or bypassing hysteresis of the phase transitions. 
All these three challenges relate to the direction of the applied electric field. 
First, it has been reported that depending on the field direction, transition temperatures may increase or decrease.\cite{Marathe.2017, li_effect_2020,bell_phenomenologically_2001} The field direction thus influences the temperature range with large ECE. Particularly, at the ferroelectric to ferroelectric transition, even the sign of the ECE depends on the electric field direction.\cite{ponomareva_bridging_2012, Marathe.2017, li_effect_2020} However, a general understanding of how the field direction influences the transition temperature of different materials is still missing. Second, unlike single crystals, the grains in polycrystals are subject to different and also low-symmetric field directions. To better understand the field direction dependence of ECE, careful sampling of the applied electric field direction is still in need. Third, it has been reported that the reduction in thermal hysteresis with increasing electric field strength depends on the direction of the applied field.\cite{Marathe.2017,li_effects_2022} 
To the best of our knowledge, the influence of the electric field direction on thermal hysteresis has only been studied for high symmetric directions (i.e.,\ $[100]$, $[110]$, and $[111]$), and while the transition temperature and ECE have been analyzed also on the $(001)$, ($\bar{1}10$), and (0$\bar{1}$1) planes, the knowledge of the direction dependence is still incomplete for both quantities.\cite{Grunebohm.2021, Marathe.2018,Marathe.2017,  li_effects_2022,li_effect_2020,ma_tailoring_2018, Liu.2016, bai_electrocaloric_2012, Moya.2014,Novak.2013,bell_phenomenologically_2001}

To address these gaps in knowledge, we reveal the impact of the direction of an applied electrical field on ferroelectric transition temperatures, thermal hysteresis, and electrocaloric response in the prototypical ferroelectric BaTiO$_3$. 
We find that although low-symmetric electric field directions cannot enhance the maximal ECE, they can reduce thermal hysteresis and widen the temperature window where large ECE appears. The full analysis of all possible field directions suggested a texturing condition where large conventional ECE can be expected at room temperature. Furthermore, we introduce a simple descriptor for the dependence of ferroelectric transition temperatures on the direction of an applied electric field for ferroelectric materials in general. 
}

\section{Method}

Our main method is the effective Hamiltonian by Zhong {\it{et al.}}\cite{Zhong.1994,Zhong.1995} parametrized with density functional theory (DFT) calculations for BaTiO$_3$\cite{Nishimatsu.2010} and PbTiO$_3$.\cite{nishimatsu_molecular_2012} \ch{In this coarse-grained approach} the energy surface is expressed as a function of local optical displacement vectors, i.e.\ dipole moments, and local acoustic displacement vectors, i.e.\ local strains, and a global homogeneous strain. \ch{Due to the coarse-graining and the additional internal optimization of strain, instead of the 15 atomistic degrees of five atoms, only the 3 degrees of freedom of the local dipole moment are explicitly considered in each unit cell. \cite{nishimatsu_direct_2013} }
This method has been successful in predicting ferroelectric phase diagrams and functional responses\cite{grunebohm_influence_2023, walizer_finite-temperature_2006} and we use it to characterize phase transitions\ch{ and directly simulate the electrocaloric} response. For these purposes, we perform molecular dynamics (MD) simulations using the open-source {\it{feram}} code\cite{Nishimatsu.2008} together with a simulation box with 36$\times$36$\times$36 unit cells (side length: 14.35~nm), periodic boundary conditions and the Nos\'e-Poincar\'e thermostat\cite{Bond.1999} with a time step of 1~fs.

\begin{figure}[t]
    \centering
    \includegraphics[width=0.5\textwidth]{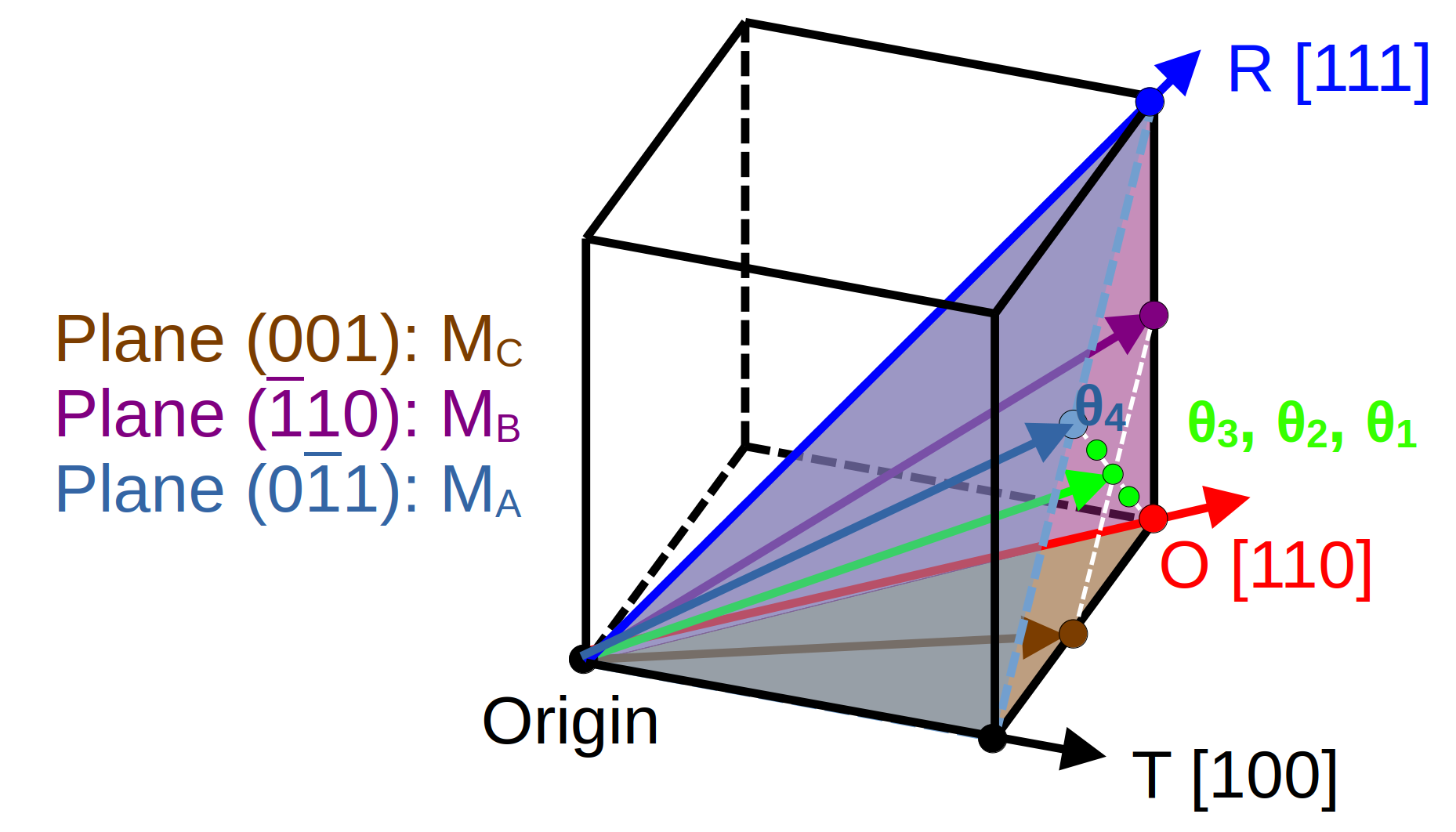}
    \caption{Sketch of sampled electric field directions. In addition to the \ch{$[100]$, $[110]$, and $[111]$ directions parallel to the spontaneous polarization in tetragonal (T), orthogonal (O), and rhombohedral (R) phases, respectively, we sample low symmetric directions on the $(001)$, $(\bar{1}10)$, and $(0\bar{1}1)$ planes, which correspond respectively to the polarization direction of the monoclinic phases M$_\text{A}$, M$_\text{B}$, and M$_\text{C}$, as well as the directions $\theta_{1}-\theta_{4}$ along $(0.75, 0.67, 0.083)$, $(0.78, 0.60, 0.18)$, $(0.81, 0.48, 0.33)$, and $(0.82, 0.41, 0.41)$.}}
    \label{fig:setup}
\end{figure}

Due to the symmetry of the \ch{perovksite structure}, the full space of electric field directions can be sampled by those directions 
 illustrated in Fig.~\ref{fig:setup}:\cite{vanderbilt_monoclinic_2001}  
The $(001)$ plane with field directions \ch{between $[100]$ and $[110]$, the ($\bar{1}10$) plane between $[110]$ and $[111]$, the (0$\bar{1}$1) plane between $[100]$ and $[111]$}, and the low symmetric directions in the region bounded by these planes. For the latter, we sample three directions $\theta_{1-3}$ on the connection line between $[110]$ and $\theta_4$ which is on the (0$\bar{1}$1) plane.

To model zero field cooling (ZFC) and electric field cooling (FC), we \ch{randomly initialize the dipoles at 380~K, i.e.\ in the paralectric phase, either without an electric field or under an external electric field of 100~kV/cm, respectively. 
Then we cool down to 25~K.} Starting from these pre-converged configurations at 25~K, zero field heating (ZFH) or field heating (FH) simulations are performed. 
\ch{The properties of interest are sampled at least every 5~K. Only around the upper FE-FE transition, we sample every  1~K. }
At each temperature, we thermalize the system for 60~ps and average for 20~ps.

To classify the symmetry of the FE phases in the applied electric field, the notation by Vanderbilt and Cohen\cite{vanderbilt_monoclinic_2001} based on the three Cartesian components of~$\vec{P}$ ($a > b > c > 0$) is followed: $\langle$a,0,0$\rangle$ for
tetragonal phases (T); $\langle$a,a,0$\rangle$ for orthorhombic phases (O); $\langle$a,a,a$\rangle$ for rhombohedral phases (R);  $\langle$b,b,a$\rangle$, $\langle$a,a,b$\rangle$, $\langle$a,b,0$\rangle$ for the monoclinic phases M$_\text{A}$, M$_\text{B}$, and M$_\text{C}$, respectively, and $\langle$a,b,c$\rangle$ for triclinic phases (Tri). 
\ch{Furthermore, we follow the convention from Ref.~\onlinecite{rose_giant_2012} and determine the transition temperatures by the maximal change in one chosen polarization component with temperature. Note that this definition gives a well-defined point on the so-called Widom line also above the critical field strength of the transition. For simplicity, we use the terms \ch{ $T_C^{ZFC}$, $T_C^{FC}$, $T_C^{ZFH}$, $T_C^{FH}$, for both, the transition temperatures and for the corresponding points on the Widom line, for zero-field cooling, field cooling, zero-field heating and field heating, respectively.}} Two repetitive simulations are done for the three high symmetric electric field directions and the repetitive error of these temperatures is within 5~K.

For completeness, the ferroelectric field hysteresis of BaTiO$_3$ is also recorded at 200~K under an electric field along $[100]$, which is the direction of the spontaneous polarization, under one pointing 15$^{\circ}$ away from $[100]$ on the $(001)$ plane, and along the direction $\theta_1$. The predicted coercive fields for these cases are  265~kV/cm, 96~kV/cm, and 49~kV/cm, respectively. Furthermore, the coercive fields for the $[100]$ direction are 75~kV/cm for BaTiO$_3$ at 50~K and 
230~kV/cm for PbTiO$_3$ at 670~K.
\ch{Note that as we simulate an ideal, homogeneous material without surfaces or defects that could act as nucleation centers, the simulated coercive fields, transition field strengths $\vec{E}_t$, critical field strengths of phase transitions, and thermal hysteresis are larger than their experimental values due to the so-called Landauer's paradox.\cite{landauer_electrostatic_1957, Grunebohm.2021} For example, the coercive field strength of unstrained single crystal BaTiO$_3$ at room temperature is around 1~kV/cm.\cite{choi_enhancement_2004} For another example, a critical field strength of 10~kV/cm and 40~kV/cm has been found for BaTiO$_3$ experimentally\cite{Novak.2013b} and in simulations,\cite{Marathe.2017, durdiev_determining_2024} respectively, under the electrical field along $[100]$. Furthermore, the field-induced changes in transition temperatures reported in experiments may be also influenced by the microstructure.}

\ch{The ECE is directly determined using the following protocol: Firstly, we equilibrate the thermalized FC configurations in the microcanonical ensemble for 80~ps and measure the actual initial temperature (${T_{\text{in}}}$) by averaging over another 40~ps. Secondly, we slowly ramp down the electric field \ch{still under adiabatic conditions} with a ramping rate of 1~kV/(cm$\cdot$ps). 
Lastly, we equilibrate the system for 80~ps without an external electric field and measure the final temperature by averaging over 100~ps, \ch{see also Appendix Fig.~\ref{fig:partialcancel}}. 
Note that the direction of the polarization may change during the field ramping, e.g., at $\vec{E}_t$, but no switching of the polarization direction at the coercive field is possible for the chosen field protocol, cf.~Fig.~\ref{fig:overviewsketchECE}.} 
Due to the coarse-graining, the specific heat in our model is underestimated by a factor of 5, and thus we correspondingly rescale the $\Delta T_{\text{ad.}}$ by 1/5.\cite{nishimatsu_direct_2013} 

\ch{According to Eqs.~\eqref{eq:TD_P} and \eqref{eq:TD_CC}, $\Delta T_{\text{ad.}}$ depends linearly on temperature.\cite{Zhang.2021, Marathe.2018,Grunebohm.2018}
To make fair comparisons between the actual adiabatic temperature changes and the underlying changes of entropy, we thus report both $\Delta T_{\text{ad.}}$ and the normalized ECE ($\Delta T_{\text{ad.}} / T_{{\text{in}}}$). Nonetheless, we would like to stress that we focus not on the exact values but on the trends of transition temperatures, thermal hysteresis, and ECE under electric fields along various directions.}

Supplementary, we \ch{introduce a simple phenomenological descriptor based on Landau's theory for} the field direction dependence of the ferroelectric phase transition temperatures.
In Landau's theory, the energy term that explicitly shows the coupling between the polarization and an external electric field ($\vec{E}$) is given as $\vec{E} \cdot \vec{P}$. As the transition temperature ($T_C$) between two phases depends on their relative energies, 
one may thus expect that the transition temperature depends on $\vec{E}$ as
\begin{equation} \label{eq:landau} T_C (\vec{E}) \sim \vec{E} \cdot \Delta \vec{P}\;,\end{equation}
with $\Delta \vec{P}$ being the polarization difference between both phases. \ch{For simplicity, we approximate $\Delta \vec{P}$ to the spontaneous polarization jump at phase transitions in ZFC simulations and we only consider the variants of the phases with the smallest angle to the applied field.}

\section{Results}

\subsection{Phase diagrams}\label{results_A}
\ch{During cooling, BaTiO$_3$ shows three first-order phase transitions between cubic (C), tetragonal (T), orthorhombic (O), and rhombohedral (R) phases,}
while PbTiO$_3$ only shows the C$\leftrightarrow$T transition. \ch{For simplicity, we also refer to the transitions by their temperature sequence from high to low temperature as the first (PE-FE), second, and third transition.} \ch{The effective Hamiltonian predicts the polarization changes at these transitions during ZFC as ($+28$,$0$,$0$),  ($-8$,$+28$,$0$), and ($-5$,$-5$,$+25$) for BaTiO$_3$; and ($+35$,$0$,$0$) for PbTiO$_3$, in units of $\mu \text{C/cm}^2$ . \ch{These values are in good agreement with previous reports based on a different method.\cite{herchig_2015_electrocaloric}}

The zero field cooling (heating) transition temperatures, $T_{C}^{\mathrm{ZFC}}$ ($T_{C}^{\mathrm{ZFH}}$), in the MD simulations are  280~K (295~K), 140~K (185~K), and 75~K (130~K) for BaTiO$_3$; and $595$~K for PbTiO$_3$.}
\ch{These values underestimate experimental values of 393~K (397~K), 278~K (285~K), and 183~K (200~K) for BaTiO$_3$\cite{aksel_advances_2010} and 765~K for PbTiO$_3$,\cite{wu_pressure-induced_2005}} due to the DFT calculations used for the parametrization of the model as well as neglected anharmonicities, and other phonon modes.\cite{Nishimatsu.2010} \ch{Note that due to the explicit temperature dependence of the ECE, cf.\ Eqs.~\eqref{eq:TD_P} and \eqref{eq:TD_CC}, we may thus underestimate the maximal caloric response quantitatively.}

\begin{figure}[h!]
    \centering
    \includegraphics[width=0.5\textwidth, keepaspectratio]{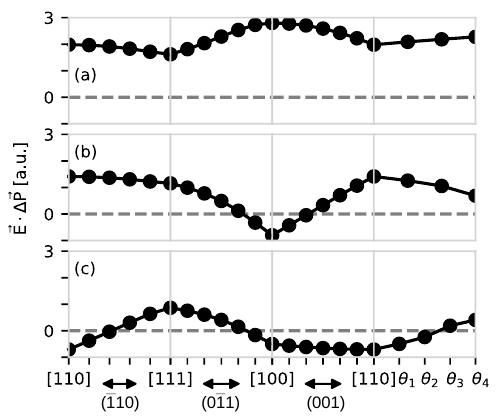}
    \caption{\ch{Direction dependence of the Landau's theory based descriptor for transition temperatures ($\vec{E} \cdot \Delta \vec{P}$) at the (a) first, (b) second, and (c) third phase transitions of BaTiO$_3$. $\Delta \vec{P}$ is approximated to the spontaneous polarization jump at zero field cooling.} }
    \label{fig:EdotP}
\end{figure}

\begin{figure*}
    \begin{center}
    \includegraphics[width=1\textwidth]{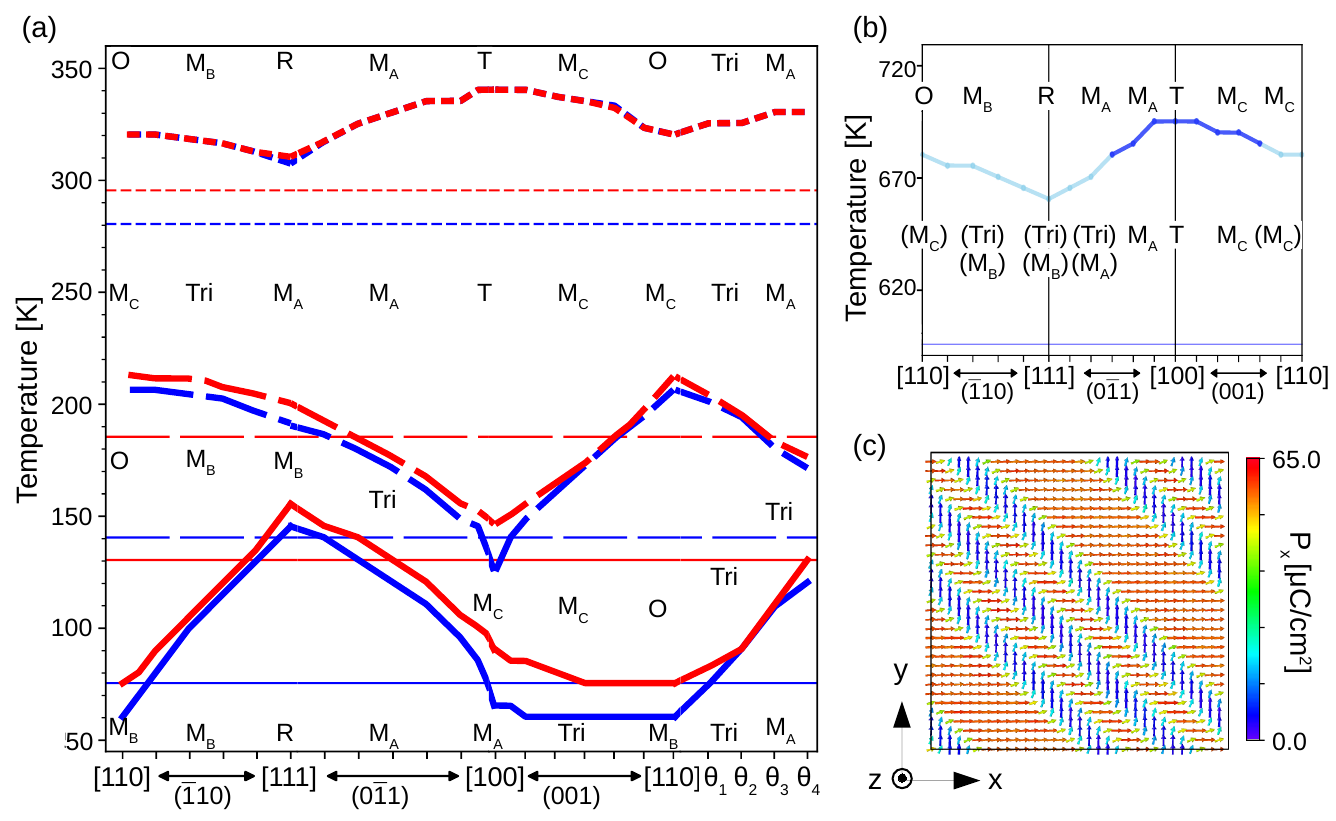}
    \caption{
Temperature-electric field direction phase diagram of (a) BaTiO$_3$ and (b) PbTiO$_3$ in an electric field of 100~kV/cm. The phases are annotated for each high symmetric direction and each plane of low symmetric direction. Thick blue and red curves \ch{mark $T_C^{\mathrm{FC}}$ and $T_C^{\mathrm{FH}}$, the transition temperatures or points on the Widom line in the external field, and thin horizontal blue and red lines} mark $T_C^{\mathrm{ZFC}}$ and $T_C^{\mathrm{ZFH}}$, the transition temperatures without field. \ch{Line styles distinguish first, second, and third transition}.  
In~(b), dark and light blue curves mark \ch{$T_C^{\mathrm{FC}}$ of }single- and multi-domain states, respectively. Possible macroscopic symmetries of the observed multidomain structures are shown in parenthesis. (c) \ch{Example of a multidomain} structure induced in PbTiO$_3$ by a field along $[110]$ at 640~K. Each arrow represents the time-averaged (20~ps) local polarization and is color-coded according to its component along the $x$-axis ($P_x$).
}
    \label{fig:BaTiO$_3$_PbTiO$_3$}
    \end{center}
\end{figure*}

How do the transition temperatures change with the direction of the applied field?
\ch{Figure~\ref{fig:EdotP} shows the qualitative changes of the transition temperatures as predicted by Eq.~\eqref{eq:landau} with zero-field polarization data, only.}
\ch{At the first transition shown in Fig.~\ref{fig:EdotP}~(a), the descriptor is always positive and shows a maximum, a local minimum, and the global minimum for field directions $[100]$, $[110]$, and $[111]$, respectively. Thus, an increase of the transition temperature, i.e.\ $T_C^{\mathrm{FC}}>T_C^{\mathrm{ZFC}}$, is expected for all field directions with a maximal increase of $T_C^{\mathrm{FC}}$ in a field along $[100]$ which parallel to the spontaneous polarization of the tetragonal phase.}
Compared with BaTiO$_3$, the larger $\Delta \vec P$ for PbTiO$_3$ suggests a larger increase in this PE-FE transition temperature.

At the second transition of BaTiO$_3$, \ch{an increase of transition temperature is predicted for most directions, see Fig.~\ref{fig:EdotP}~(b). However, close to $[100]$ the projection of polarization on the field direction is larger in the high-temperature phase than in the low-temperature one, and Eq.~\eqref{eq:landau} suggests a decrease in the transition temperature under the field.}
\ch{At the third transition shown in Fig.~\ref{fig:EdotP}~(c), $\vec{E} \cdot \Delta \vec{P}$ is positive if the field points close to $[111]$, i.e.,\ if the field stabilizes the R phase. 
For other directions, particularly on the $(001)$-plane, the high-temperature phase is stabilized by the field.}

Qualitatively, all these predictions are in good agreement with $T_C^{\mathrm{FC}}$ found directly in MD simulations as shown by thick blue curves in Fig.~\ref{fig:BaTiO$_3$_PbTiO$_3$} for a field strength of 100~kV/cm and in Appendix Fig.~\ref{fig:phasediagram_BaTiO$_3$_10kVcm} for a field strength of 10~kV/cm, which are also in good agreement with literature\cite{li_effect_2020, Marathe.2017}. Firstly, indeed, \ch{the predicted increase or decrease of the transition temperatures for high-symmetric field directions are well reproduced with 
} $T_{C}^{\mathrm{FC}}>T_{C}^{\mathrm{ZFC}}$ for all field directions at the first transition, for fields not pointing close to $[100]$ at the second transition, and for fields on the $(001)$ plane at the third transition.
Secondly, it is true that \ch{the induced increase of $T_{C}^{\mathrm{FC}}$ is larger in PbTiO$_3$ compared to BaTiO$_3$}. \ch{This shows that the projection of $\Delta \vec{P}$ in ZFC on the direction of the applied field is indeed a good descriptor to predict the direction-dependence of the field-induced changes in ferroelectric transition temperatures in general.} 

However, the simple estimate given by Eq.~\eqref{eq:landau} does neither include field-induced changes in polarization, nor the domain structure, nor the dynamics of the transitions, and thus fails to quantitatively predict \ch{the field directions with the cross-over from enhanced to reduced transition temperatures $T_{C}^{ZFC} = T_{C}^{FC}$.} 
For example, the simple descriptor overestimates the direction range with $T_C^{\mathrm{FC}}<T_C^{\mathrm{ZFC}}$ at the second transition around $[100]$ by about 33\%, and underestimates the range with $T_C^{\mathrm{FC}}>T_C^{\mathrm{ZFC}}$ at the third transition around $[111]$  by about 68\%. \footnote{At the second transition of BaTiO$_3$, Eq.~\eqref{eq:landau}
predicts that the field directions where $T_C^{\mathrm{FC}}=T_C^{\mathrm{ZFC}}$ are along 22$^{\circ}$ away from $[100]$ on the $(0\bar11)$ plane and along 16$^{\circ}$ away from $[100]$ on the $(001)$ plane, while MD predicts smaller angles: 8$^{\circ}$ and 5$^{\circ}$.
At the third transition, Eq.~\eqref{eq:landau} predicts that the field directions where $T_C^{\mathrm{FC}}=T_C^{\mathrm{ZFC}}$ are 19$^{\circ}$ away from $[111]$ on the $(\bar110)$ plane and within 35$^{\circ}$ away from $[111]$ on the $(0\bar11)$ plane, while MD predicts larger angles: 28$^{\circ}$ and 51$^{\circ}$.}
\ch{Furthermore,} the differences between BaTiO$_3$ and PbTiO$_3$ are not \ch{quantitatively correct}: \ch{While $\Delta \vec{P}$ of BaTiO$_3$ and PbTiO$_3$ differs by 25\%, the actual increase in $T_C^{\mathrm{FC}}$ of PbTiO$_3$ is larger than this for all field directions, e.g.,\ 116\% for the field along $[111]$.}

\ch{Microscopically, the coupling between the polarization and external electric field shows important differences in BaTiO$_3$ and PbTiO$_3$, see Fig.~\ref{fig:BaTiO$_3$_PbTiO$_3$}.}
\ch{If the field is applied parallel to the spontaneous polarization of a phase, it changes the ordering and the magnitude of the dipoles in both materials. For fields pointing to all other directions, the symmetry of the phases also changes. In the paraelectric phase at high temperatures, the direction of the field fully determines the symmetry of the phase, e.g.,\ the M$_\text{B}$ phase is induced for fields on the $(\bar110)$ plane.
Within a ferroelectric phase, it's more complex:} Rotating the field away from the direction of the spontaneous polarization reduces the symmetry to monoclinic or triclinic. \ch{In BaTiO$_3$ a slight homogeneous polarization rotation is induced while the systems stay in a single domain state. An example of the local dipole distributions is shown in Appendix Fig.~\ref{fig:Pdist}~(a).}

For PbTiO$_3$, one has to distinguish two regimes \ch{of field directions}:  \ch{The first regime close to [100] is marked in dark blue in Fig.~\ref{fig:BaTiO$_3$_PbTiO$_3$}~(b). There, also PbTiO$_3$ remains single domain. However the polarization rotation is smaller compared to BaTiO$_3$, see also Appendix Fig.~\ref{fig:Pdist}~(b)}.
\ch{In the second regime indicated by light blue in Fig.~\ref{fig:BaTiO$_3$_PbTiO$_3$}~(b), PbTiO$_3$ decomposes into multidomain states. For example, for the field direction $[110]$, tetragonal domains with polarization pointing along $[100]$ and $[010]$ occur which are separated by charge-neutral T 90$^{\circ}$ walls, see Fig.~\ref{fig:BaTiO$_3$_PbTiO$_3$}~(c).} \ch{The larger tendency of PbTiO$_3$ to form domains is in agreement to previous observations without external field} and has been  related to the larger polarization-strain coupling of PbTiO$_3$.\cite{nishimatsu_molecular_2012}
Note that even though a field along $[111]$ would equally stabilize all three types of $\langle 100 \rangle$-domains, we always only find \ch{domains with two polarization directions.} Replicate simulations show that  the width and polarization direction of the domains \ch{and thus the macroscopic symmetry of the ferroelectric phase depend on} nucleation.  After domains \ch{with two polarization directions have been} nucleated, the formation of \ch{domains along the third direction} would result in wall crossings and \ch{were} too high in energy in our simulations. 
Importantly, neither the appearance of multidomain structures nor wall orientation and spacing influence the field dependence of $T_C^{\mathrm{FC}}$ at the given temperature resolution. In addition, the differences between $T_C^{\mathrm{FC}}$ for fields along $[100]$ and $[111]$ are similar for PbTiO$_3$ (35~K) and BaTiO$_3$ (33~K).

\ch{Next, we discuss the influence of the field direction on thermal hysteresis for BaTiO$_3$.}
In general, thermal hysteresis decreases with the increasing strength of the applied field. 
At the first transition, the critical field strengths are smaller than 100~kV/cm for fields along $[100]$ and $[110]$.\cite{li_effects_2022,Marathe.2017, durdiev_determining_2024} 
In agreement, we observe continuous changes in polarization without thermal hysteresis under fields of 100~kV/cm with the chosen temperature resolution of 5~K for all directions, except for $[111]$ where a clear polarization jump and thermal hysteresis of 5~K are observed at the transition.
The transitions are at the same time isostructural for most field directions, except for fields pointing on the $(\bar110)$ plane. Although our resolution is not sufficient for a conclusive analysis, one may speculate that the critical field is larger around $[111]$  as 
dP/dT is less smooth for these field directions compared to the directions with isostrucutral transitions.

At the FE-FE transitions, three different scenarios are possible for the transition temperatures: First, the field induces a decrease in both transition temperatures, $T_C^{\mathrm{FC}}< T_C^{\mathrm{ZFC}}$ and $T_C^{\mathrm{FH}}< T_C^{\mathrm{ZFH}}$, if it stabilizes the high-temperature phase, e.g.\ at the second transition under fields along $[100]$ and at the third transition under fields on the $(001)$ plane. Here, electric fields induce a larger reduction in $T_C^{\mathrm{FH}}$ than in $T_C^{\mathrm{FC}}$.
Second, an increase in both transition temperatures, $T_C^{\mathrm{FC}}> T_C^{\mathrm{ZFC}}$ and $T_C^{\mathrm{FH}}> T_C^{\mathrm{ZFH}}$, is induced by the field, if it stabilizes the low-temperature phase, e.g.\ at the second transition under fields along $(\bar110)$ and at the third transition under fields along $[111]$. Here, the field-induced increase is larger in $T_C^{\mathrm{FC}}$ than in $T_C^{\mathrm{FH}}$.
Third, the combination of $T_C^{\mathrm{FH}}< T_C^{\mathrm{ZFH}}$ and $T_C^{\mathrm{FC}}> T_C^{\mathrm{ZFC}}$, which is guaranteed to reduce the thermal hysteresis, is induced by fields along low symmetric directions.

\begin{figure}[t]
    \centering
    \includegraphics[width=0.5\textwidth, keepaspectratio]{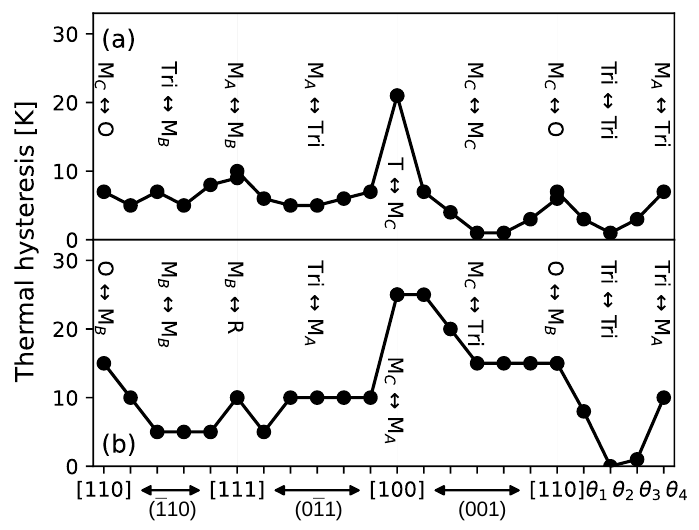}
    \caption{Field direction dependence of thermal hysteresis in BaTiO$_3$ under fields of 100~kV/cm at the (a) second and (b) third transition. \ch{For comparison, the thermal hysteresis without field is 45~K and 55~K at the second and third transitions, respectively.}}
    \label{fig:thermalhysteresis}
\end{figure}

The resulting thermal hysteresis for both FE-FE transitions is shown in Fig.~\ref{fig:thermalhysteresis}.\footnote{Note that the simulated hysteresis without fields is about 45~K and 55~K at the second and third transition, which are about 8 and 4 times larger than the experimental values, respectively, as expected for the idealized defect-free material.\cite{samara_1971_pressure}} 
For all high symmetric directions,\ch{ a finite thermal hysteresis remains for the used field strength of 100~kV/cm,} in agreement with predicted larger critical field strengths for these transitions.~\cite{Marathe.2017, li_effects_2022}\ch{\footnote{\ch{For the field along $[110]$ critical field strengths of 125~kV/cm and $>$~200~kV/cm have been reported for the second and third transitions, respectively.\cite{li_effects_2022}}}} 
Maximal thermal hysteresis persists in the field exactly along $[100]$ for both FE-FE transitions. 
\ch{Already} small rotations away from this direction result in abrupt reductions in thermal hysteresis at the second transition. At the third transition, the reduction in thermal hysteresis is most obvious if the rotation is on the $(0\bar11)$ plane. These changes with the field direction are related to abrupt changes of  
 $T_C^{FC}$ with the change in symmetry of the high-temperature phase, whereas $T_C^{FH}$ is less field direction dependent around $[100]$.
Generally, a rotation of the field direction that changes only the symmetry of the high-temperature (low-temperature) phase causes an abrupt change in $T_C^{\mathrm{FC}}$ ($T_C^{\mathrm{FH}}$), while \ch{the other transition temperature} changes linearly with the field rotation.

Low symmetric directions reduce thermal hysteresis if they decrease and increase the transition temperatures under cooling and heating, respectively. At the second transition, hysteresis is reduced to around 5~K for fields on the $(\bar110)$ and $(0\bar11)$ planes. The hysteresis is further reduced, if the transition becomes isostructural, e.g.,\ Tri$\leftrightarrow$Tri or M$_\text{C}$$\leftrightarrow$M$_\text{C}$ transitions under fields along $\theta_2$ or 
on the $(001)$ plane, respectively. At the third transition, thermal hysteresis can be reduced to around 15~K and 10~K for fields on the $(001)$ and $(0\bar11)$ planes\ch{, respectively; or to 5~K or lower at the M$_\text{B}$$\leftrightarrow$M$_\text{B}$ transition for fields on the $(\bar110)$ plane or at the Tri$\leftrightarrow$Tri transition for fields along $\theta_i$, respectively. In summary, thermal hysteresis is the smallest under isostructural transition, and the maximal value for fields along $[100]$ is halved if the field direction is slightly modified.}

\subsection{Electrocaloric response}

\begin{figure*}[t]
    \centering    
    \includegraphics[width=0.7\textwidth, keepaspectratio]{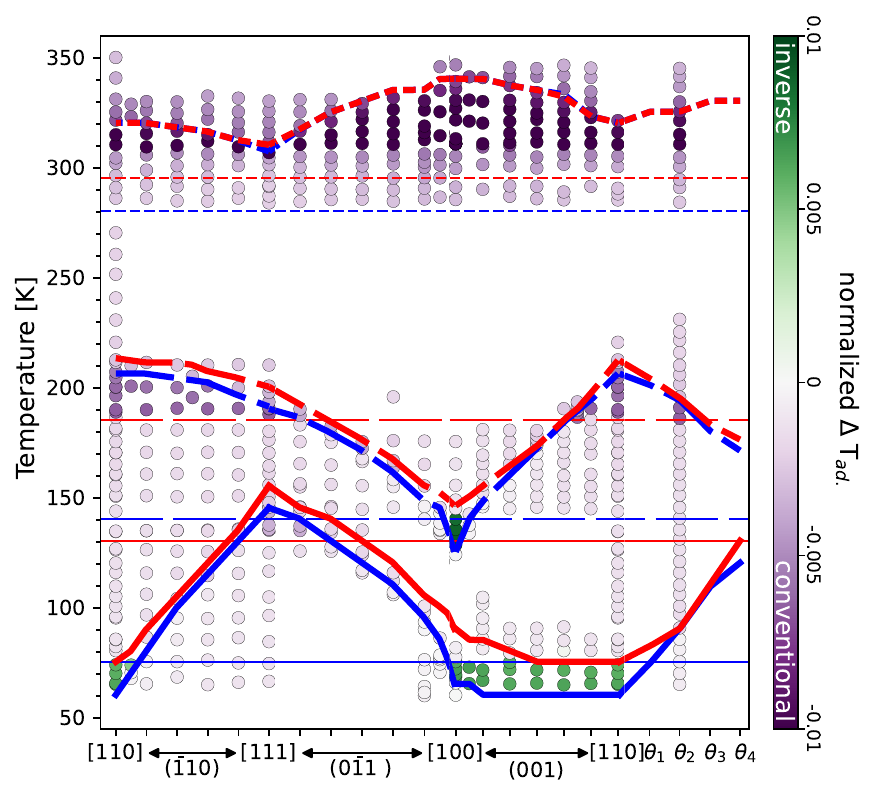}
    \caption{Temperature and electric field direction dependence of the normalized ECE ($\Delta T_{\text{ad.}}/T_{\text{in}}$) during field removal after field cooling of  BaTiO$_3$. \ch{Thick blue and red curves \ch{mark $T_C^{\mathrm{FC}}$ and $T_C^{\mathrm{FH}}$, and thin horizontal blue and red lines} mark $T_C^{\mathrm{ZFC}}$ and $T_C^{\mathrm{ZFH}}$. The first, second, and third transitions are distinguished by the line style}.  \ch{Dots mark temperatures and field directions for which the ECE has been directly determined and are colored by the 
     normalized ECE. Large conventional/inverse ECEs are observed between $T_C^{\mathrm{FC}}$ and $T_C^{\mathrm{ZFH}}$/$T_C^{\mathrm{ZFC}}$.} 
    }
    \label{fig:ECE}
\end{figure*}

\begin{figure}[t]
    \centering    
    \includegraphics[width=0.48\textwidth, keepaspectratio]{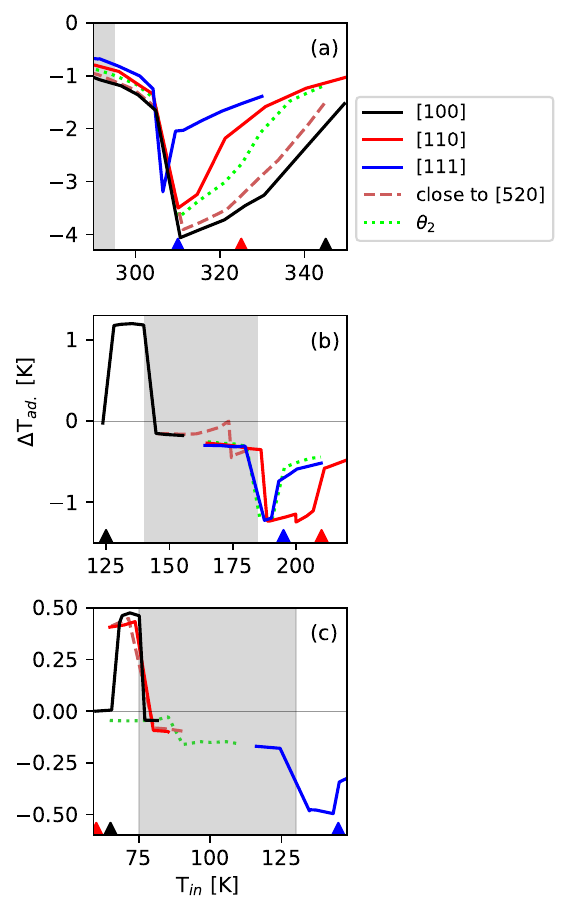}
    \caption{Temperature dependence of the ECE of BTO for representative electric field directions at (a) the first\ch{, (b) second, and (c) third} phase transitions. The color code of directions is consistent with Fig.~\ref{fig:setup}. The triangles mark the $T_C^{\mathrm{FC}}$, \ch{i.e.\ temperature with the maximal slope with polarization,} under fields along the corresponding directions. Gray regions are the corresponding ZF coexistence ranges. \ch{For simplicity, only the large temperature changes near transitions are shown.}}
    \label{fig:everything}
\end{figure}

Which electric field direction yields a large reversible ECE in a broad temperature range and is thus optimal for applications?
To answer this question, we first analyze the adiabatic temperature change for representative field directions. The normalized adiabatic temperature change ($\Delta T/T_{\text{in}}$) of the pre-cooled system is shown in Fig.~\ref{fig:ECE} as colored dots at the sampled temperatures $T_{\text{in}}$ and field directions. Fig.~\ref{fig:everything} provides a closer look at the temperature dependence of $\Delta T_{\text{ad.}}$ for representative field directions.

At all transitions, a large conventional ECE is restricted to temperatures where the system undergoes a phase transition from the low- to high-temperature phase during field removal, e.g.\ from T to C. This is possible only in the temperature windows between $T_C^{\mathrm{ZFH}}$ and $T_C^{\mathrm{FC}}$. 
The responses within the FE phases are one order of magnitude smaller compared to those at phase transitions, i.e.\ $|\Delta T_{{\text{ad.}}}/{T_{\text{in}}}|\leq$ 0.005 above the PE-FE transition, and  $\leq$ 0.002 near the FE-FE lower transitions, see lightly colored dots in Fig.~\ref{fig:ECE}. Consequently, the ECE peaks show sharp shoulders at each boundary to an FE phase, as shown in Fig.~\ref{fig:everything}.
\ch{The maximal $\Delta T_{{\text{ad.}}}$ ($\Delta T_{{\text{ad.}}}/T_{\text{in}}$)  at the first, second, and third transitions are $-$4~K ($-$0.013), $-$1.2~K ($-$0.006), and $-$0.5~K ($-$0.004), respectively. } 

At the PE-FE transition, i.e.\ around 310~K, the conventional ECE is maximal, even if rescaled with temperature. At this transition, the largest temperature change is induced at the lowest temperature where the system still undergoes the phase transition, i.e.,\ at 310~K, and $\Delta T_{\text{ad.}}$ decreases continuously with increasing temperature, consistent with the literature.\cite{li_effect_2020, Marathe.2017} The only exception is the $[111]$ direction. Here, one can distinguish between the contribution to $\Delta T_{\text{ad.}}$ from the continuous entropy change, which is at most 2~K at high temperatures, and the contribution from the latent heat at the first-order phase transition, which has a peak of 3~K at below $T_C^{\mathrm{FC}}$. This distinction is no longer possible for the stronger field of 200~kV/cm as it has been reported before.\cite{li_effect_2020}

The maximal ECE at the PE-FE transition can be induced by a field along the $[100]$ direction, \ch{see black line in Fig.~\ref{fig:everything}~(a).} Any rotation of the field causes a gradual reduction in the coupling between the polarization and the applied electric field and in the magnitude of the ECE, e.g.\ $\Delta T_{\text{ad.}}$ for a field along $[110]$ is only around 88\% of that for a field along $[100]$. In addition, as a consequence of the discussed direction dependence of $T_C^{\mathrm{FC}}$, the temperature window where large ECE appears is also strongly influenced by field direction. This temperature window is the widest for a field along $[100]$ at the PE-FE transition and it narrows to less than half for a field along $[110]$, as shown in Fig.~\ref{fig:ECE} and Fig.~\ref{fig:everything}~(a).

At the FE-FE transitions, the field direction dependence of the temperature window where large responses occur is still strong. However, the temperature and the applied field direction have only a minor impact on the magnitude of the caloric response, as long as the phase transition takes place during field removal, see Fig.~\ref{fig:everything}~(b)--(c). This observation is consistent with the literature.\cite{Marathe.2017} 
At the second transition, the maximal conventional ECE is around 1~K and the widest temperature window where large responses appear is around 25~K under a field along $[110]$. This temperature window varies only slightly for fields between $[110]$ and up to 75$^{\circ}$ away from $[110]$ on the ${\bar110}$ plane. While this temperature window narrows drastically when the applied field deviates from $[110]$ on the $(001)$ plane, as shown in Fig.~\ref{fig:ECE}.

An inverse ECE is possible at the FE-FE transitions for field-cooled samples 
if $T_C^{\mathrm{FC}}$ is below $T_C^{\mathrm{ZFC}}$, marked as green dots in Fig.~\ref{fig:ECE}. For the chosen field strength, the maximal inverse ECE at the second transition is $+$1.2~K ($\Delta T_{{\text{ad.}}}/T_{\text{in}}$: $+$0.009) with a temperature window of 10~K under a field along $[100]$, and the maximal inverse ECE at the third transition is $+0.5$~K ($\Delta T_{{\text{ad.}}}/T_{\text{in}}$: $+$0.007) with a temperature window of 15~K under fields on the $(001)$ plane, see Fig.~\ref{fig:everything}~(b) and~(c).
The inverse ECE at the second transition is very sensitive to electric field direction: a field deviates merely 8$^{\circ}$ away from $[100]$ can turn the strong inverse response into a weak conventional one.
The inverse ECE at the third transition is less sensitive to field direction: these inverse responses vanish for fields that deviate from the $(001)$ plane.

As discussed, at the FE-FE phase transition, as long as the phase transition takes place during field removal, the magnitude of the ECE does not depend much on the field direction. The maximal field direction dependence of the $\Delta T_{\text{ad.}}$ peaks is about 10\% and is found for the inverse ECE at the third transition for fields between $[100]$ and $[110]$, see Fig.~\ref{fig:everything}~(c).
\ch{This direction dependence of inverse ECE is related to the direction dependence of $\vec{E}_t$ in combination with the small conventional ECE within the M$_\text{C}$ phase, see Appendix Fig.~\ref{fig:partialcancel}:  For the field along $[100]$, the 
M$_\text{C}$$\leftrightarrow$M$_\text{A}$ transition takes place at around 40~kV/cm, while for the other two field directions, the M$_\text{C}$$\leftrightarrow$Tri transition occurs at around 10~kV/cm. The ramping of the field from 100 to 10~kV/cm induces a larger conventional ECE in the initial phase compared to the ramping from 100 to 40~kV/cm and thus the remaining overall inverse ECE is larger for the latter field direction.}

So far, we have discussed the ECE of field-cooled samples during field removal, for which a large conventional response is possible between $T_C^{\mathrm{ZFH}}$ and $T_C^{\mathrm{FC}}$, i.e.,\ between red lines and thick blue curves in Fig.~\ref{fig:ECE}, by the transition from low- to high-temperature phase. In principle, a phase transition and a large conventional response would also be induced between $T_C^{\mathrm{ZFH}}$ and $T_C^{\mathrm{FH}}$, i.e.,\ between red lines and thick red curves in Fig.~\ref{fig:ECE}, in field-heated samples. However, in the coexistence range between $T_C^{\mathrm{FH}}$ and $T_C^{\mathrm{FC}}$, this larger response is not reversible, and thus cannot be used in a device. 
Strictly speaking, also the inverse ECE discussed so far is induced between $T_C^{\mathrm{FC}}$ and $T_C^{\mathrm{ZFC}}$ and thus not reversible in a cycling field.
A large, reversible, inverse ECE is possible only between $T_C^{\mathrm{FH}}$ and $T_C^{\mathrm{ZFC}}$, with $T_C^{\mathrm{FH}}< T_C^{\mathrm{ZFC}}$.\cite{Marathe.2018}
While this condition is not yet fulfilled for 100~kV/cm, $T_C^{\mathrm{FH}}$ decreases with the field strength, both for fields close to [100] at the second transition an for field on the  $(001)$-plane for the third transition. For example,  the field-induced decrease in the latter case is 15~K for 10~kV/cm, see Appendix Fig.~\ref{fig:phasediagram_BaTiO$_3$_10kVcm}, and 55~K for 100~kV/cm. As $T_C^{\mathrm{FH}}\approx T_C^{\mathrm{ZFC}}$ for the chosen field strength, a reversibe inverse ECE has to be expected for slightly larger field strengths both, at the second transition for the field along 100 and for the third transition, if the field is applied on the (001)-plane.

What change in caloric responses can be expected for misaligned fields \ch{in single crystals or for (un-)}textured ceramics? At the first transition, $\Delta T_{\text{ad.}}$ larger than 1~K is observed above 310~K under arbitrarily pointing fields, as shown in Fig.~\ref{fig:ECE} and Fig.~\ref{fig:everything}~(a). As discussed, the maximal responses for all fields occur around 310~K but the temperature windows where large responses occur vary. Textured ceramics where most grains experience a field along $[100]$ are expected to exhibit the largest overall $\Delta T_{\text{ad.}}$ with the widest temperature window. The overall $\Delta T_{\text{ad.}}$ and the temperature window shrink with increasing volume fraction of grains that experience a field along $[111]$.
At the second transition at temperatures between 185 and 210~K, for fields between $[110]$ and up to 75$^{\circ}$ away from $[110]$ on the $(\bar110)$ plane, and fields between $[110]$ and $\theta_2$, which is around 15$^{\circ}$ away from $[110]$, the magnitude of the conventional ECE during field removal in single crystals is almost independent of the field direction, as shown in Fig.~\ref{fig:everything}~(b). This conventional ECE is expected in textured ceramics where most grains experience fields along $[110]$ or on the $(\bar110)$ plane. If the ceramics are untextured or textured in an unfavorable direction, e.g.\ $[100]$, then the overall caloric responses and the temperature windows where large responses appear are expected to shrink.
In summary, misaligned fields and textured ceramics where most grains experience fields on the $\{\bar110\}$ plane are promising for reliable, conventional caloric responses at the first and second transitions, respectively. 
Similarly, at the third transition, an inverse ECE is likely to be observed in textured ceramics where most grains experience fields on the $\{001\}$ plane during field removal.
On the other hand, between 130 and 150~K in our simulation, the caloric response is very sensitive to field direction and thus strictly textured ceramics are required. As shown in Fig.~\ref{fig:ECE}, $\Delta T_{\text{ad.}}$ in single crystals is very sensitive to field direction at the second transition under fields around $[100]$ and at the third transition under fields around $[111]$ at temperatures. Thus, the large inverse or conventional responses at these temperatures are only expected in textured ceramics where most grains experience fields along $[100]$ or $[111]$, respectively. Otherwise, only a small $\Delta T_{\text{ad.}}$ is expected, as the inverse and conventional responses from grains experiencing fields along $[111]$ and $[100]$ partly compensate each other and the responses from other grains are very weak.

In summary, a reversible, conventional ECE of more than 1~K can be expected at temperatures larger than 305~K and temperatures between 185--206~K for fields on the $(\bar110)$ plane as well as those between $[110]$ and $\theta_2$ under a field strength of 100~kV/cm. 
\ch{Additionally, for fields along $\theta_2$, this large conventional ECE is accompanied by almost zero thermal hysteresis, and thus a high coefficient of performance is expected.}

\section{Summary and Conclusions}
Ideal ECE materials for solid-state cooling technologies should show large and reversible caloric responses at the desired working temperatures. Using coarse-grained molecular dynamics simulations, we have revealed the field-direction dependence of ferroelectric transition temperatures, thermal hysteresis, temperature windows where large responses appear, and the magnitude of the caloric responses. At the ferroelectric to ferroelectric transitions, even the sign of the field-induced change in transition temperature depends on the field direction, which could partly explain the broad range of their experimentally reported values for (un)textured ceramics. We found that although low-symmetric electric field directions cannot enhance the maximal ECE, they can reduce thermal hysteresis and widen the temperature window where large ECE appears, and thus increase the thermodynamic cooling efficiency.

Below we summarize the texture conditions for ceramics and the expected caloric responses and reversibility from high- to low-temperature in reality.
At temperatures close to the zero-field heating transition temperature ($T_C^{\mathrm{ZFH}}$) at the paraelectric to ferroelectric transition, which corresponds to around 398~K in reality, BaTiO$_3$ shows the largest conventional ECE under fields along $[100]$ with no thermal hysteresis for fields stronger than the critical field strength. 
Fields along the $\langle111\rangle$ direction can reduce the total caloric response and narrow the temperature range where large responses occur in ceramics and should be avoided by proper texturing. As this caloric response takes place at temperatures much higher than room temperature, further treatments, e.g.\ inclusion\cite{lv_manipulation_2020,uddin_effect_2021,bratton_phase_1967,lisenkov_tuning_2018}, is needed to lower the transition temperature for room-temperature cooling technologies.

At temperatures close to $T_C^{\mathrm{ZFH}}$ at the T--O transition, corresponding to around 285~K in reality, a large and reversible conventional response is expected in textured ceramics where most grains experience fields on the $\{\bar110\}$ plane. In agreement, a reversible conventional ECE of 1.33~K has been found at 288~K under a field along $\langle 011 \rangle$\cite{li_near-room-temperature_2021}. Ceramics textured correspondingly, which can be experimentally prepared by e.g.,\ templated grain growth,\cite{meier_highly_2016, schultheis_ferroelectric_2023} are expected to be suitable for room-temperature applications.

At temperatures close to the zero-field cooling transition temperature ($T_C^{\mathrm{ZFC}}$) at the T--O transition, corresponding to around 278~K in reality, only strictly textured ceramics where almost all grains experience fields along $[100]$ might exhibit large inverse ECE. Similarly, at temperatures close to $T_C^{\mathrm{ZFH}}$ at the O--R transition, corresponding to around 200~K, only strictly textured ceramics where almost grains experience fields along $[111]$ show large conventional ECE. These two caloric responses could occur at similar temperatures under a strong electric field and ceramics need to be carefully textured to avoid partial cancellation of the inverse and conventional ECEs.

At temperatures close to $T_C^{\mathrm{ZFC}}$ at the O--R transition, corresponding to around 183~K in reality, inverse ECEs are expected in textured ceramics where most grains experience fields on the $\{001\}$ plane. However, the reversible responses are only expected under strong fields that can drive transition temperature below $T_C^{\mathrm{ZFC}}$.

Supplementarily, we have analyzed ferroelectric transition temperatures for all possible directions of the applied electric field on two perovskite oxides -- BaTiO$_3$ and PbTiO$_3$. Even though BaTiO$_3$ is susceptible to polarization rotation and PbTiO$_3$ shows multidomain structures under non-collinear electric fields, the changes in the transition temperatures with field direction can be predicted qualitatively based on the spontaneous polarization jumps at the phase transitions of the single-domain material. This simple descriptor allows future screening and optimization of transition temperatures in ferroelectric materials.

\begin{acknowledgments}
All authors acknowledge support from the Deutsche Forschungsgemeinschaft (DFG), Germany. 
F.W. is grateful for the financial support under grant GRK2495/K, L.H. and A.G. under grant GR 4792/3.

\end{acknowledgments}

\bibliography{main}

\begin{thebibliography}{42}%
\makeatletter
\providecommand \@ifxundefined [1]{%
 \@ifx{#1\undefined}
}%
\providecommand \@ifnum [1]{%
 \ifnum #1\expandafter \@firstoftwo
 \else \expandafter \@secondoftwo
 \fi
}%
\providecommand \@ifx [1]{%
 \ifx #1\expandafter \@firstoftwo
 \else \expandafter \@secondoftwo
 \fi
}%
\providecommand \natexlab [1]{#1}%
\providecommand \enquote  [1]{``#1''}%
\providecommand \bibnamefont  [1]{#1}%
\providecommand \bibfnamefont [1]{#1}%
\providecommand \citenamefont [1]{#1}%
\providecommand \href@noop [0]{\@secondoftwo}%
\providecommand \href [0]{\begingroup \@sanitize@url \@href}%
\providecommand \@href[1]{\@@startlink{#1}\@@href}%
\providecommand \@@href[1]{\endgroup#1\@@endlink}%
\providecommand \@sanitize@url [0]{\catcode `\\12\catcode `\$12\catcode
  `\&12\catcode `\#12\catcode `\^12\catcode `\_12\catcode `\%12\relax}%
\providecommand \@@startlink[1]{}%
\providecommand \@@endlink[0]{}%
\providecommand \url  [0]{\begingroup\@sanitize@url \@url }%
\providecommand \@url [1]{\endgroup\@href {#1}{\urlprefix }}%
\providecommand \urlprefix  [0]{URL }%
\providecommand \Eprint [0]{\href }%
\providecommand \doibase [0]{http://dx.doi.org/}%
\providecommand \selectlanguage [0]{\@gobble}%
\providecommand \bibinfo  [0]{\@secondoftwo}%
\providecommand \bibfield  [0]{\@secondoftwo}%
\providecommand \translation [1]{[#1]}%
\providecommand \BibitemOpen [0]{}%
\providecommand \bibitemStop [0]{}%
\providecommand \bibitemNoStop [0]{.\EOS\space}%
\providecommand \EOS [0]{\spacefactor3000\relax}%
\providecommand \BibitemShut  [1]{\csname bibitem#1\endcsname}%
\let\auto@bib@innerbib\@empty
\bibitem [{\citenamefont {Ožbolt}\ \emph {et~al.}(2014)\citenamefont
  {Ožbolt}, \citenamefont {Kitanovski}, \citenamefont {Tušek},\ and\
  \citenamefont {Poredoš}}]{ozbolt_electrocaloric_2014}%
  \BibitemOpen
  \bibfield  {author} {\bibinfo {author} {\bibfnamefont {M.}~\bibnamefont
  {Ožbolt}}, \bibinfo {author} {\bibfnamefont {A.}~\bibnamefont {Kitanovski}},
  \bibinfo {author} {\bibfnamefont {J.}~\bibnamefont {Tušek}}, \ and\ \bibinfo
  {author} {\bibfnamefont {A.}~\bibnamefont {Poredoš}},\ }\href {\doibase
  10.1016/j.ijrefrig.2013.07.001} {\bibfield  {journal} {\bibinfo  {journal}
  {IJR}\ }\bibinfo {series} {New {Developments} in {Magnetic}
  {Refrigeration}},\ \textbf {\bibinfo {volume} {37}},\ \bibinfo {pages} {16}
  (\bibinfo {year} {2014})}\BibitemShut {NoStop}%
\bibitem [{\citenamefont {Aprea}\ \emph {et~al.}(2017)\citenamefont {Aprea},
  \citenamefont {Greco}, \citenamefont {Maiorino},\ and\ \citenamefont
  {Masselli}}]{aprea_comparison_2017}%
  \BibitemOpen
  \bibfield  {author} {\bibinfo {author} {\bibfnamefont {C.}~\bibnamefont
  {Aprea}}, \bibinfo {author} {\bibfnamefont {A.}~\bibnamefont {Greco}},
  \bibinfo {author} {\bibfnamefont {A.}~\bibnamefont {Maiorino}}, \ and\
  \bibinfo {author} {\bibfnamefont {C.}~\bibnamefont {Masselli}},\ }\href
  {\doibase 10.18280/ijht.350130} {\bibfield  {journal} {\bibinfo  {journal}
  {IJHT}\ }\textbf {\bibinfo {volume} {35}},\ \bibinfo {pages} {225} (\bibinfo
  {year} {2017})}\BibitemShut {NoStop}%
\bibitem [{\citenamefont {Shi}\ \emph {et~al.}(2019)\citenamefont {Shi},
  \citenamefont {Han}, \citenamefont {Li}, \citenamefont {Yang}, \citenamefont
  {Lu}, \citenamefont {Zhong}, \citenamefont {Chen}, \citenamefont {Zhang},\
  and\ \citenamefont {Qian}}]{shi_electrocaloric_2019}%
  \BibitemOpen
  \bibfield  {author} {\bibinfo {author} {\bibfnamefont {J.}~\bibnamefont
  {Shi}}, \bibinfo {author} {\bibfnamefont {D.}~\bibnamefont {Han}}, \bibinfo
  {author} {\bibfnamefont {Z.}~\bibnamefont {Li}}, \bibinfo {author}
  {\bibfnamefont {L.}~\bibnamefont {Yang}}, \bibinfo {author} {\bibfnamefont
  {S.-G.}\ \bibnamefont {Lu}}, \bibinfo {author} {\bibfnamefont
  {Z.}~\bibnamefont {Zhong}}, \bibinfo {author} {\bibfnamefont
  {J.}~\bibnamefont {Chen}}, \bibinfo {author} {\bibfnamefont {Q.~M.}\
  \bibnamefont {Zhang}}, \ and\ \bibinfo {author} {\bibfnamefont
  {X.}~\bibnamefont {Qian}},\ }\href {\doibase 10.1016/j.joule.2019.03.021}
  {\bibfield  {journal} {\bibinfo  {journal} {Joule}\ }\textbf {\bibinfo
  {volume} {3}},\ \bibinfo {pages} {1200} (\bibinfo {year} {2019})}\BibitemShut
  {NoStop}%
\bibitem [{\citenamefont {Wang}\ \emph {et~al.}(2020)\citenamefont {Wang},
  \citenamefont {Zhang}, \citenamefont {Usui}, \citenamefont {Benedict},
  \citenamefont {Hirose}, \citenamefont {Lee}, \citenamefont {Kalb},\ and\
  \citenamefont {Schwartz}}]{wang_high-performance_2020}%
  \BibitemOpen
  \bibfield  {author} {\bibinfo {author} {\bibfnamefont {Y.}~\bibnamefont
  {Wang}}, \bibinfo {author} {\bibfnamefont {Z.}~\bibnamefont {Zhang}},
  \bibinfo {author} {\bibfnamefont {T.}~\bibnamefont {Usui}}, \bibinfo {author}
  {\bibfnamefont {M.}~\bibnamefont {Benedict}}, \bibinfo {author}
  {\bibfnamefont {S.}~\bibnamefont {Hirose}}, \bibinfo {author} {\bibfnamefont
  {J.}~\bibnamefont {Lee}}, \bibinfo {author} {\bibfnamefont {J.}~\bibnamefont
  {Kalb}}, \ and\ \bibinfo {author} {\bibfnamefont {D.}~\bibnamefont
  {Schwartz}},\ }\href {\doibase 10.1126/science.aba2648} {\bibfield  {journal}
  {\bibinfo  {journal} {Science}\ }\textbf {\bibinfo {volume} {370}},\ \bibinfo
  {pages} {129} (\bibinfo {year} {2020})}\BibitemShut {NoStop}%
\bibitem [{\citenamefont {Greco}\ and\ \citenamefont
  {Masselli}(2020)}]{greco_electrocaloric_2020}%
  \BibitemOpen
  \bibfield  {author} {\bibinfo {author} {\bibfnamefont {A.}~\bibnamefont
  {Greco}}\ and\ \bibinfo {author} {\bibfnamefont {C.}~\bibnamefont
  {Masselli}},\ }\href {\doibase 10.3390/magnetochemistry6040067} {\bibfield
  {journal} {\bibinfo  {journal} {Magnetochemistry}\ }\textbf {\bibinfo
  {volume} {6}},\ \bibinfo {pages} {67} (\bibinfo {year} {2020})}\BibitemShut
  {NoStop}%
\bibitem [{\citenamefont {Torell{\'o}}\ and\ \citenamefont
  {Defay}(2022)}]{Torello.2022}%
  \BibitemOpen
  \bibfield  {author} {\bibinfo {author} {\bibfnamefont {A.}~\bibnamefont
  {Torell{\'o}}}\ and\ \bibinfo {author} {\bibfnamefont {E.}~\bibnamefont
  {Defay}},\ }\href {\doibase 10.1002/aelm.202101031} {\bibfield  {journal}
  {\bibinfo  {journal} {Adv. Electron. Mater.}\ }\textbf {\bibinfo {volume}
  {8}},\ \bibinfo {pages} {2101031} (\bibinfo {year} {2022})}\BibitemShut
  {NoStop}%
\bibitem [{\citenamefont {Marathe}\ \emph {et~al.}(2017)\citenamefont
  {Marathe}, \citenamefont {Renggli}, \citenamefont {Sanlialp}, \citenamefont
  {Karabasov}, \citenamefont {Shvartsman}, \citenamefont {Lupascu},
  \citenamefont {Gr{\"u}nebohm},\ and\ \citenamefont {Ederer}}]{Marathe.2017}%
  \BibitemOpen
  \bibfield  {author} {\bibinfo {author} {\bibfnamefont {M.}~\bibnamefont
  {Marathe}}, \bibinfo {author} {\bibfnamefont {D.}~\bibnamefont {Renggli}},
  \bibinfo {author} {\bibfnamefont {M.}~\bibnamefont {Sanlialp}}, \bibinfo
  {author} {\bibfnamefont {M.~O.}\ \bibnamefont {Karabasov}}, \bibinfo {author}
  {\bibfnamefont {V.~V.}\ \bibnamefont {Shvartsman}}, \bibinfo {author}
  {\bibfnamefont {D.~C.}\ \bibnamefont {Lupascu}}, \bibinfo {author}
  {\bibfnamefont {A.}~\bibnamefont {Gr{\"u}nebohm}}, \ and\ \bibinfo {author}
  {\bibfnamefont {C.}~\bibnamefont {Ederer}},\ }\href {\doibase
  10.1103/PhysRevB.96.014102} {\bibfield  {journal} {\bibinfo  {journal} {Phys.
  Rev. B}\ }\textbf {\bibinfo {volume} {96}},\ \bibinfo {pages} {014102}
  (\bibinfo {year} {2017})}\BibitemShut {NoStop}%
\bibitem [{\citenamefont {Moya}\ \emph {et~al.}(2014)\citenamefont {Moya},
  \citenamefont {Kar-Narayan},\ and\ \citenamefont {Mathur}}]{Moya.2014}%
  \BibitemOpen
  \bibfield  {author} {\bibinfo {author} {\bibfnamefont {X.}~\bibnamefont
  {Moya}}, \bibinfo {author} {\bibfnamefont {S.}~\bibnamefont {Kar-Narayan}}, \
  and\ \bibinfo {author} {\bibfnamefont {N.~D.}\ \bibnamefont {Mathur}},\
  }\href {\doibase 10.1038/nmat3951} {\bibfield  {journal} {\bibinfo  {journal}
  {Nat. Mater.}\ }\textbf {\bibinfo {volume} {13}},\ \bibinfo {pages} {439}
  (\bibinfo {year} {2014})}\BibitemShut {NoStop}%
\bibitem [{\citenamefont {Mischenko}\ \emph {et~al.}(2006)\citenamefont
  {Mischenko}, \citenamefont {Zhang}, \citenamefont {Scott}, \citenamefont
  {Whatmore},\ and\ \citenamefont {Mathur}}]{Mischenko.2006}%
  \BibitemOpen
  \bibfield  {author} {\bibinfo {author} {\bibfnamefont {A.~S.}\ \bibnamefont
  {Mischenko}}, \bibinfo {author} {\bibfnamefont {Q.}~\bibnamefont {Zhang}},
  \bibinfo {author} {\bibfnamefont {J.~F.}\ \bibnamefont {Scott}}, \bibinfo
  {author} {\bibfnamefont {R.~W.}\ \bibnamefont {Whatmore}}, \ and\ \bibinfo
  {author} {\bibfnamefont {N.~D.}\ \bibnamefont {Mathur}},\ }\href {\doibase
  10.1126/science.1123811} {\bibfield  {journal} {\bibinfo  {journal}
  {Science}\ }\textbf {\bibinfo {volume} {311}},\ \bibinfo {pages} {1270}
  (\bibinfo {year} {2006})}\BibitemShut {NoStop}%
\bibitem [{\citenamefont {Neese}\ \emph {et~al.}(2008)\citenamefont {Neese},
  \citenamefont {Chu}, \citenamefont {Lu}, \citenamefont {Wang}, \citenamefont
  {Furman},\ and\ \citenamefont {Zhang}}]{Neese.2008}%
  \BibitemOpen
  \bibfield  {author} {\bibinfo {author} {\bibfnamefont {B.}~\bibnamefont
  {Neese}}, \bibinfo {author} {\bibfnamefont {B.}~\bibnamefont {Chu}}, \bibinfo
  {author} {\bibfnamefont {S.-G.}\ \bibnamefont {Lu}}, \bibinfo {author}
  {\bibfnamefont {Y.}~\bibnamefont {Wang}}, \bibinfo {author} {\bibfnamefont
  {E.}~\bibnamefont {Furman}}, \ and\ \bibinfo {author} {\bibfnamefont {Q.~M.}\
  \bibnamefont {Zhang}},\ }\href {\doibase 10.1126/science.1159655} {\bibfield
  {journal} {\bibinfo  {journal} {Science}\ }\textbf {\bibinfo {volume}
  {321}},\ \bibinfo {pages} {821} (\bibinfo {year} {2008})}\BibitemShut
  {NoStop}%
\bibitem [{\citenamefont {Wu}\ and\ \citenamefont
  {Cohen}(2017{\natexlab{a}})}]{Wu.2017}%
  \BibitemOpen
  \bibfield  {author} {\bibinfo {author} {\bibfnamefont {H.~H.}\ \bibnamefont
  {Wu}}\ and\ \bibinfo {author} {\bibfnamefont {R.~E.}\ \bibnamefont {Cohen}},\
  }\href {\doibase 10.1088/1361-648X/aa94db} {\bibfield  {journal} {\bibinfo
  {journal} {J. Phys.: Condens. Matter}\ }\textbf {\bibinfo {volume} {29}},\
  \bibinfo {pages} {485704} (\bibinfo {year} {2017}{\natexlab{a}})}\BibitemShut
  {NoStop}%
\bibitem [{\citenamefont {Gr{\"u}nebohm}\ \emph {et~al.}(2018)\citenamefont
  {Gr{\"u}nebohm}, \citenamefont {Ma}, \citenamefont {Marathe}, \citenamefont
  {Xu}, \citenamefont {Albe}, \citenamefont {Kalcher}, \citenamefont {Meyer},
  \citenamefont {Shvartsman}, \citenamefont {Lupascu},\ and\ \citenamefont
  {Ederer}}]{Grunebohm.2018}%
  \BibitemOpen
  \bibfield  {author} {\bibinfo {author} {\bibfnamefont {A.}~\bibnamefont
  {Gr{\"u}nebohm}}, \bibinfo {author} {\bibfnamefont {Y.-B.}\ \bibnamefont
  {Ma}}, \bibinfo {author} {\bibfnamefont {M.}~\bibnamefont {Marathe}},
  \bibinfo {author} {\bibfnamefont {B.-X.}\ \bibnamefont {Xu}}, \bibinfo
  {author} {\bibfnamefont {K.}~\bibnamefont {Albe}}, \bibinfo {author}
  {\bibfnamefont {C.}~\bibnamefont {Kalcher}}, \bibinfo {author} {\bibfnamefont
  {K.-C.}\ \bibnamefont {Meyer}}, \bibinfo {author} {\bibfnamefont {V.~V.}\
  \bibnamefont {Shvartsman}}, \bibinfo {author} {\bibfnamefont {D.~C.}\
  \bibnamefont {Lupascu}}, \ and\ \bibinfo {author} {\bibfnamefont
  {C.}~\bibnamefont {Ederer}},\ }\href {\doibase 10.1002/ente.201800166}
  {\bibfield  {journal} {\bibinfo  {journal} {Energy Technol.}\ }\textbf
  {\bibinfo {volume} {6}},\ \bibinfo {pages} {1491} (\bibinfo {year}
  {2018})}\BibitemShut {NoStop}%
\bibitem [{\citenamefont {Moya}\ \emph {et~al.}(2013)\citenamefont {Moya},
  \citenamefont {Stern-Taulats}, \citenamefont {Crossley}, \citenamefont
  {González-Alonso}, \citenamefont {Kar-Narayan}, \citenamefont {Planes},
  \citenamefont {Mañosa},\ and\ \citenamefont {Mathur}}]{moya_giant_2013}%
  \BibitemOpen
  \bibfield  {author} {\bibinfo {author} {\bibfnamefont {X.}~\bibnamefont
  {Moya}}, \bibinfo {author} {\bibfnamefont {E.}~\bibnamefont {Stern-Taulats}},
  \bibinfo {author} {\bibfnamefont {S.}~\bibnamefont {Crossley}}, \bibinfo
  {author} {\bibfnamefont {D.}~\bibnamefont {González-Alonso}}, \bibinfo
  {author} {\bibfnamefont {S.}~\bibnamefont {Kar-Narayan}}, \bibinfo {author}
  {\bibfnamefont {A.}~\bibnamefont {Planes}}, \bibinfo {author} {\bibfnamefont
  {L.}~\bibnamefont {Mañosa}}, \ and\ \bibinfo {author} {\bibfnamefont
  {N.~D.}\ \bibnamefont {Mathur}},\ }\href {\doibase 10.1002/adma.201203823}
  {\bibfield  {journal} {\bibinfo  {journal} {Adv Mater}\ }\textbf {\bibinfo
  {volume} {25}},\ \bibinfo {pages} {1360} (\bibinfo {year}
  {2013})}\BibitemShut {NoStop}%
\bibitem [{\citenamefont {Rose}\ and\ \citenamefont
  {Cohen}(2012)}]{rose_giant_2012}%
  \BibitemOpen
  \bibfield  {author} {\bibinfo {author} {\bibfnamefont {M.~C.}\ \bibnamefont
  {Rose}}\ and\ \bibinfo {author} {\bibfnamefont {R.~E.}\ \bibnamefont
  {Cohen}},\ }\href {\doibase 10.1103/PhysRevLett.109.187604} {\bibfield
  {journal} {\bibinfo  {journal} {Phys. Rev. Lett.}\ }\textbf {\bibinfo
  {volume} {109}},\ \bibinfo {pages} {187604} (\bibinfo {year}
  {2012})}\BibitemShut {NoStop}%
\bibitem [{\citenamefont {Lisenkov}\ and\ \citenamefont
  {Ponomareva}(2018)}]{lisenkov_tuning_2018}%
  \BibitemOpen
  \bibfield  {author} {\bibinfo {author} {\bibfnamefont {S.}~\bibnamefont
  {Lisenkov}}\ and\ \bibinfo {author} {\bibfnamefont {I.}~\bibnamefont
  {Ponomareva}},\ }\href {\doibase 10.1103/PhysRevMaterials.2.055402}
  {\bibfield  {journal} {\bibinfo  {journal} {Phys. Rev. Mater.}\ }\textbf
  {\bibinfo {volume} {2}},\ \bibinfo {pages} {055402} (\bibinfo {year}
  {2018})}\BibitemShut {NoStop}%
\bibitem [{\citenamefont {Wu}\ and\ \citenamefont
  {Cohen}(2017{\natexlab{b}})}]{wu_electric-field-induced_2017}%
  \BibitemOpen
  \bibfield  {author} {\bibinfo {author} {\bibfnamefont {H.~H.}\ \bibnamefont
  {Wu}}\ and\ \bibinfo {author} {\bibfnamefont {R.~E.}\ \bibnamefont {Cohen}},\
  }\href {\doibase 10.1103/PhysRevB.96.054116} {\bibfield  {journal} {\bibinfo
  {journal} {Phys. Rev. B}\ }\textbf {\bibinfo {volume} {96}},\ \bibinfo
  {pages} {054116} (\bibinfo {year} {2017}{\natexlab{b}})}\BibitemShut
  {NoStop}%
\bibitem [{\citenamefont {Gr{\"u}nebohm}\ and\ \citenamefont
  {Nishimatsu}(2018)}]{grunebohm_optimizing_2018}%
  \BibitemOpen
  \bibfield  {author} {\bibinfo {author} {\bibfnamefont {A.}~\bibnamefont
  {Gr{\"u}nebohm}}\ and\ \bibinfo {author} {\bibfnamefont {N.}~\bibnamefont
  {Nishimatsu}, \bibfnamefont {T}},\ }\href@noop {} {\bibfield  {journal}
  {\bibinfo  {journal} {8\textsuperscript{th} International Conference on
  Caloric Cooling (Thermag VIII). Proceedings: Darmstadt, Germany, September
  16-20, 2018.}\ } (\bibinfo {year} {2018})}\BibitemShut {NoStop}%
\bibitem [{\citenamefont {Taxil}\ \emph {et~al.}(2022)\citenamefont {Taxil},
  \citenamefont {Lallart}, \citenamefont {Ducharne}, \citenamefont {Nguyen},
  \citenamefont {Kuwano}, \citenamefont {Ono},\ and\ \citenamefont
  {Sebald}}]{taxil_modeling_2022}%
  \BibitemOpen
  \bibfield  {author} {\bibinfo {author} {\bibfnamefont {G.}~\bibnamefont
  {Taxil}}, \bibinfo {author} {\bibfnamefont {M.}~\bibnamefont {Lallart}},
  \bibinfo {author} {\bibfnamefont {B.}~\bibnamefont {Ducharne}}, \bibinfo
  {author} {\bibfnamefont {T.~T.}\ \bibnamefont {Nguyen}}, \bibinfo {author}
  {\bibfnamefont {H.}~\bibnamefont {Kuwano}}, \bibinfo {author} {\bibfnamefont
  {T.}~\bibnamefont {Ono}}, \ and\ \bibinfo {author} {\bibfnamefont
  {G.}~\bibnamefont {Sebald}},\ }\href {\doibase 10.1063/5.0107429} {\bibfield
  {journal} {\bibinfo  {journal} {PRL}\ }\textbf {\bibinfo {volume} {132}},\
  \bibinfo {pages} {144101} (\bibinfo {year} {2022})}\BibitemShut {NoStop}%
\bibitem [{\citenamefont {Li}\ \emph {et~al.}(2020)\citenamefont {Li},
  \citenamefont {Li}, \citenamefont {Wu}, \citenamefont {Li}, \citenamefont
  {Wang}, \citenamefont {Qin}, \citenamefont {Su}, \citenamefont {Qiao},
  \citenamefont {Guo},\ and\ \citenamefont {Bai}}]{li_effect_2020}%
  \BibitemOpen
  \bibfield  {author} {\bibinfo {author} {\bibfnamefont {Z.}~\bibnamefont
  {Li}}, \bibinfo {author} {\bibfnamefont {J.}~\bibnamefont {Li}}, \bibinfo
  {author} {\bibfnamefont {H.-H.}\ \bibnamefont {Wu}}, \bibinfo {author}
  {\bibfnamefont {J.}~\bibnamefont {Li}}, \bibinfo {author} {\bibfnamefont
  {S.}~\bibnamefont {Wang}}, \bibinfo {author} {\bibfnamefont {S.}~\bibnamefont
  {Qin}}, \bibinfo {author} {\bibfnamefont {Y.}~\bibnamefont {Su}}, \bibinfo
  {author} {\bibfnamefont {L.}~\bibnamefont {Qiao}}, \bibinfo {author}
  {\bibfnamefont {D.}~\bibnamefont {Guo}}, \ and\ \bibinfo {author}
  {\bibfnamefont {Y.}~\bibnamefont {Bai}},\ }\href {\doibase
  10.1016/j.actamat.2020.03.020} {\bibfield  {journal} {\bibinfo  {journal}
  {Acta Mater.}\ }\textbf {\bibinfo {volume} {191}},\ \bibinfo {pages} {13}
  (\bibinfo {year} {2020})}\BibitemShut {NoStop}%
\bibitem [{\citenamefont {Li}\ \emph {et~al.}(2022)\citenamefont {Li},
  \citenamefont {Wu}, \citenamefont {Li}, \citenamefont {Wang}, \citenamefont
  {Qin}, \citenamefont {He}, \citenamefont {Liu}, \citenamefont {Su},
  \citenamefont {Qiao}, \citenamefont {Lookman},\ and\ \citenamefont
  {Bai}}]{li_effects_2022}%
  \BibitemOpen
  \bibfield  {author} {\bibinfo {author} {\bibfnamefont {Z.}~\bibnamefont
  {Li}}, \bibinfo {author} {\bibfnamefont {H.-H.}\ \bibnamefont {Wu}}, \bibinfo
  {author} {\bibfnamefont {J.}~\bibnamefont {Li}}, \bibinfo {author}
  {\bibfnamefont {S.}~\bibnamefont {Wang}}, \bibinfo {author} {\bibfnamefont
  {S.}~\bibnamefont {Qin}}, \bibinfo {author} {\bibfnamefont {J.}~\bibnamefont
  {He}}, \bibinfo {author} {\bibfnamefont {C.}~\bibnamefont {Liu}}, \bibinfo
  {author} {\bibfnamefont {Y.}~\bibnamefont {Su}}, \bibinfo {author}
  {\bibfnamefont {L.}~\bibnamefont {Qiao}}, \bibinfo {author} {\bibfnamefont
  {T.}~\bibnamefont {Lookman}}, \ and\ \bibinfo {author} {\bibfnamefont
  {Y.}~\bibnamefont {Bai}},\ }\href {\doibase 10.1016/j.actamat.2022.117784}
  {\bibfield  {journal} {\bibinfo  {journal} {Acta Mater.}\ }\textbf {\bibinfo
  {volume} {228}},\ \bibinfo {pages} {117784} (\bibinfo {year}
  {2022})}\BibitemShut {NoStop}%
\bibitem [{\citenamefont {Marathe}\ \emph {et~al.}(2018)\citenamefont
  {Marathe}, \citenamefont {Ederer},\ and\ \citenamefont
  {Gr{\"u}nebohm}}]{Marathe.2018}%
  \BibitemOpen
  \bibfield  {author} {\bibinfo {author} {\bibfnamefont {M.}~\bibnamefont
  {Marathe}}, \bibinfo {author} {\bibfnamefont {C.}~\bibnamefont {Ederer}}, \
  and\ \bibinfo {author} {\bibfnamefont {A.}~\bibnamefont {Gr{\"u}nebohm}},\
  }\href {\doibase 10.1002/pssb.201700308} {\bibfield  {journal} {\bibinfo
  {journal} {PSS (b)}\ }\textbf {\bibinfo {volume} {255}},\ \bibinfo {pages}
  {1700308} (\bibinfo {year} {2018})}\BibitemShut {NoStop}%
\bibitem [{\citenamefont {Ma}\ \emph {et~al.}(2018)\citenamefont {Ma},
  \citenamefont {Xu}, \citenamefont {Albe},\ and\ \citenamefont
  {Grünebohm}}]{ma_tailoring_2018}%
  \BibitemOpen
  \bibfield  {author} {\bibinfo {author} {\bibfnamefont {Y.-B.}\ \bibnamefont
  {Ma}}, \bibinfo {author} {\bibfnamefont {B.-X.}\ \bibnamefont {Xu}}, \bibinfo
  {author} {\bibfnamefont {K.}~\bibnamefont {Albe}}, \ and\ \bibinfo {author}
  {\bibfnamefont {A.}~\bibnamefont {Grünebohm}},\ }\href {\doibase
  10.1103/PhysRevApplied.10.024048} {\bibfield  {journal} {\bibinfo  {journal}
  {Phys. Rev. Appl.}\ }\textbf {\bibinfo {volume} {10}},\ \bibinfo {pages}
  {024048} (\bibinfo {year} {2018})}\BibitemShut {NoStop}%
\bibitem [{\citenamefont {Liu}\ \emph {et~al.}(2016{\natexlab{a}})\citenamefont
  {Liu}, \citenamefont {Scott},\ and\ \citenamefont {Dkhil}}]{Liu.2016}%
  \BibitemOpen
  \bibfield  {author} {\bibinfo {author} {\bibfnamefont {Y.}~\bibnamefont
  {Liu}}, \bibinfo {author} {\bibfnamefont {J.~F.}\ \bibnamefont {Scott}}, \
  and\ \bibinfo {author} {\bibfnamefont {B.}~\bibnamefont {Dkhil}},\ }\href
  {\doibase 10.1063/1.4958327} {\bibfield  {journal} {\bibinfo  {journal}
  {Appl. Phys. Rev.}\ }\textbf {\bibinfo {volume} {3}},\ \bibinfo {pages}
  {031102} (\bibinfo {year} {2016}{\natexlab{a}})}\BibitemShut {NoStop}%
\bibitem [{\citenamefont {Gr{\"u}nebohm}\ \emph {et~al.}(2021)\citenamefont
  {Gr{\"u}nebohm}, \citenamefont {Marathe}, \citenamefont {Khachaturyan},
  \citenamefont {Schiedung}, \citenamefont {Lupascu},\ and\ \citenamefont
  {Shvartsman}}]{Grunebohm.2021}%
  \BibitemOpen
  \bibfield  {author} {\bibinfo {author} {\bibfnamefont {A.}~\bibnamefont
  {Gr{\"u}nebohm}}, \bibinfo {author} {\bibfnamefont {M.}~\bibnamefont
  {Marathe}}, \bibinfo {author} {\bibfnamefont {R.}~\bibnamefont
  {Khachaturyan}}, \bibinfo {author} {\bibfnamefont {R.}~\bibnamefont
  {Schiedung}}, \bibinfo {author} {\bibfnamefont {D.~C.}\ \bibnamefont
  {Lupascu}}, \ and\ \bibinfo {author} {\bibfnamefont {V.~V.}\ \bibnamefont
  {Shvartsman}},\ }\href {\doibase 10.1088/1361-648X/ac3607} {\bibfield
  {journal} {\bibinfo  {journal} {J. Condens. Matter Phys.}\ }\textbf {\bibinfo
  {volume} {34}},\ \bibinfo {pages} {073002} (\bibinfo {year}
  {2021})}\BibitemShut {NoStop}%
\bibitem [{\citenamefont {Bai}\ \emph {et~al.}(2012)\citenamefont {Bai},
  \citenamefont {Ding}, \citenamefont {Zheng}, \citenamefont {Shi},
  \citenamefont {Cao},\ and\ \citenamefont {Qiao}}]{bai_electrocaloric_2012}%
  \BibitemOpen
  \bibfield  {author} {\bibinfo {author} {\bibfnamefont {Y.}~\bibnamefont
  {Bai}}, \bibinfo {author} {\bibfnamefont {K.}~\bibnamefont {Ding}}, \bibinfo
  {author} {\bibfnamefont {G.-P.}\ \bibnamefont {Zheng}}, \bibinfo {author}
  {\bibfnamefont {S.-Q.}\ \bibnamefont {Shi}}, \bibinfo {author} {\bibfnamefont
  {J.-L.}\ \bibnamefont {Cao}}, \ and\ \bibinfo {author} {\bibfnamefont
  {L.}~\bibnamefont {Qiao}},\ }\href {\doibase 10.1063/1.4732146} {\bibfield
  {journal} {\bibinfo  {journal} {AIP Adv.}\ }\textbf {\bibinfo {volume} {2}},\
  \bibinfo {pages} {022162} (\bibinfo {year} {2012})}\BibitemShut {NoStop}%
\bibitem [{\citenamefont {Novak}\ \emph {et~al.}(2013)\citenamefont {Novak},
  \citenamefont {Kutnjak},\ and\ \citenamefont {Pirc}}]{Novak.2013}%
  \BibitemOpen
  \bibfield  {author} {\bibinfo {author} {\bibfnamefont {N.}~\bibnamefont
  {Novak}}, \bibinfo {author} {\bibfnamefont {Z.}~\bibnamefont {Kutnjak}}, \
  and\ \bibinfo {author} {\bibfnamefont {R.}~\bibnamefont {Pirc}},\ }\href
  {\doibase 10.1209/0295-5075/103/47001} {\bibfield  {journal} {\bibinfo
  {journal} {EPL}\ }\textbf {\bibinfo {volume} {103}},\ \bibinfo {pages}
  {47001} (\bibinfo {year} {2013})}\BibitemShut {NoStop}%
\bibitem [{\citenamefont {Liu}\ \emph {et~al.}(2016{\natexlab{b}})\citenamefont
  {Liu}, \citenamefont {Scott},\ and\ \citenamefont {Dkhil}}]{Liu.2016b}%
  \BibitemOpen
  \bibfield  {author} {\bibinfo {author} {\bibfnamefont {Y.}~\bibnamefont
  {Liu}}, \bibinfo {author} {\bibfnamefont {J.~F.}\ \bibnamefont {Scott}}, \
  and\ \bibinfo {author} {\bibfnamefont {B.}~\bibnamefont {Dkhil}},\ }\href
  {\doibase 10.1063/1.4954056} {\bibfield  {journal} {\bibinfo  {journal} {APL
  Mater.}\ }\textbf {\bibinfo {volume} {4}},\ \bibinfo {pages} {064109}
  (\bibinfo {year} {2016}{\natexlab{b}})}\BibitemShut {NoStop}%
\bibitem [{\citenamefont {Gutfleisch}\ \emph {et~al.}(2016)\citenamefont
  {Gutfleisch}, \citenamefont {Gottschall}, \citenamefont {Fries},
  \citenamefont {Benke}, \citenamefont {Radulov}, \citenamefont {Skokov},
  \citenamefont {Wende}, \citenamefont {Gruner}, \citenamefont {Acet},
  \citenamefont {Entel},\ and\ \citenamefont {Farle}}]{Gutfleisch.2016}%
  \BibitemOpen
  \bibfield  {author} {\bibinfo {author} {\bibfnamefont {O.}~\bibnamefont
  {Gutfleisch}}, \bibinfo {author} {\bibfnamefont {T.}~\bibnamefont
  {Gottschall}}, \bibinfo {author} {\bibfnamefont {M.}~\bibnamefont {Fries}},
  \bibinfo {author} {\bibfnamefont {D.}~\bibnamefont {Benke}}, \bibinfo
  {author} {\bibfnamefont {I.}~\bibnamefont {Radulov}}, \bibinfo {author}
  {\bibfnamefont {K.~P.}\ \bibnamefont {Skokov}}, \bibinfo {author}
  {\bibfnamefont {H.}~\bibnamefont {Wende}}, \bibinfo {author} {\bibfnamefont
  {M.}~\bibnamefont {Gruner}}, \bibinfo {author} {\bibfnamefont
  {M.}~\bibnamefont {Acet}}, \bibinfo {author} {\bibfnamefont {P.}~\bibnamefont
  {Entel}}, \ and\ \bibinfo {author} {\bibfnamefont {M.}~\bibnamefont
  {Farle}},\ }\href@noop {} {\bibfield  {journal} {\bibinfo  {journal} {Philos.
  Trans. Royal Soc. A}\ }\textbf {\bibinfo {volume} {374}},\ \bibinfo {pages}
  {20150308} (\bibinfo {year} {2016})}\BibitemShut {NoStop}%
\bibitem [{\citenamefont {Li}\ \emph {et~al.}(2021)\citenamefont {Li},
  \citenamefont {Li}, \citenamefont {He}, \citenamefont {Hou}, \citenamefont
  {Su}, \citenamefont {Qiao},\ and\ \citenamefont
  {Bai}}]{li_near-room-temperature_2021}%
  \BibitemOpen
  \bibfield  {author} {\bibinfo {author} {\bibfnamefont {J.}~\bibnamefont
  {Li}}, \bibinfo {author} {\bibfnamefont {Z.}~\bibnamefont {Li}}, \bibinfo
  {author} {\bibfnamefont {J.}~\bibnamefont {He}}, \bibinfo {author}
  {\bibfnamefont {Y.}~\bibnamefont {Hou}}, \bibinfo {author} {\bibfnamefont
  {Y.}~\bibnamefont {Su}}, \bibinfo {author} {\bibfnamefont {L.}~\bibnamefont
  {Qiao}}, \ and\ \bibinfo {author} {\bibfnamefont {Y.}~\bibnamefont {Bai}},\
  }\href {\doibase 10.1002/pssr.202100251} {\bibfield  {journal} {\bibinfo
  {journal} {PSS (RRL)}\ }\textbf {\bibinfo {volume} {15}},\ \bibinfo {pages}
  {2100251} (\bibinfo {year} {2021})}\BibitemShut {NoStop}%
\bibitem [{\citenamefont {Mikhaleva}\ \emph {et~al.}(2012)\citenamefont
  {Mikhaleva}, \citenamefont {Flerov}, \citenamefont {Gorev}, \citenamefont
  {Molokeev}, \citenamefont {Cherepakhin}, \citenamefont {Kartashev},
  \citenamefont {Mikhashenok},\ and\ \citenamefont
  {Sablina}}]{mikhaleva_caloric_2012}%
  \BibitemOpen
  \bibfield  {author} {\bibinfo {author} {\bibfnamefont {E.}~\bibnamefont
  {Mikhaleva}}, \bibinfo {author} {\bibfnamefont {I.}~\bibnamefont {Flerov}},
  \bibinfo {author} {\bibfnamefont {M.}~\bibnamefont {Gorev}}, \bibinfo
  {author} {\bibfnamefont {M.}~\bibnamefont {Molokeev}}, \bibinfo {author}
  {\bibfnamefont {A.}~\bibnamefont {Cherepakhin}}, \bibinfo {author}
  {\bibfnamefont {A.}~\bibnamefont {Kartashev}}, \bibinfo {author}
  {\bibfnamefont {N.}~\bibnamefont {Mikhashenok}}, \ and\ \bibinfo {author}
  {\bibfnamefont {K.}~\bibnamefont {Sablina}},\ }\href {\doibase
  10.1134/S1063783412090181} {\bibfield  {journal} {\bibinfo  {journal} {J.
  Solid State Phys.}\ }\textbf {\bibinfo {volume} {54}},\ \bibinfo {pages}
  {1832} (\bibinfo {year} {2012})}\BibitemShut {NoStop}%
\bibitem [{\citenamefont {Zhong}\ \emph {et~al.}(1994)\citenamefont {Zhong},
  \citenamefont {Vanderbilt},\ and\ \citenamefont {Rabe}}]{Zhong.1994}%
  \BibitemOpen
  \bibfield  {author} {\bibinfo {author} {\bibfnamefont {W.}~\bibnamefont
  {Zhong}}, \bibinfo {author} {\bibfnamefont {D.}~\bibnamefont {Vanderbilt}}, \
  and\ \bibinfo {author} {\bibfnamefont {K.~M.}\ \bibnamefont {Rabe}},\ }\href
  {\doibase 10.1103/PhysRevLett.73.1861} {\bibfield  {journal} {\bibinfo
  {journal} {Phys. Rev. Lett.}\ }\textbf {\bibinfo {volume} {73}},\ \bibinfo
  {pages} {1861} (\bibinfo {year} {1994})}\BibitemShut {NoStop}%
\bibitem [{\citenamefont {Zhong}\ \emph {et~al.}(1995)\citenamefont {Zhong},
  \citenamefont {Vanderbilt},\ and\ \citenamefont {Rabe}}]{Zhong.1995}%
  \BibitemOpen
  \bibfield  {author} {\bibinfo {author} {\bibfnamefont {W.}~\bibnamefont
  {Zhong}}, \bibinfo {author} {\bibfnamefont {D.}~\bibnamefont {Vanderbilt}}, \
  and\ \bibinfo {author} {\bibfnamefont {K.~M.}\ \bibnamefont {Rabe}},\ }\href
  {\doibase 10.1103/PhysRevB.52.6301} {\bibfield  {journal} {\bibinfo
  {journal} {Physical Review B}\ }\textbf {\bibinfo {volume} {52}},\ \bibinfo
  {pages} {6301} (\bibinfo {year} {1995})}\BibitemShut {NoStop}%
\bibitem [{\citenamefont {Nishimatsu}\ \emph {et~al.}(2010)\citenamefont
  {Nishimatsu}, \citenamefont {Iwamoto}, \citenamefont {Kawazoe},\ and\
  \citenamefont {Waghmare}}]{Nishimatsu.2010}%
  \BibitemOpen
  \bibfield  {author} {\bibinfo {author} {\bibfnamefont {T.}~\bibnamefont
  {Nishimatsu}}, \bibinfo {author} {\bibfnamefont {M.}~\bibnamefont {Iwamoto}},
  \bibinfo {author} {\bibfnamefont {Y.}~\bibnamefont {Kawazoe}}, \ and\
  \bibinfo {author} {\bibfnamefont {U.~V.}\ \bibnamefont {Waghmare}},\ }\href
  {\doibase 10.1103/PhysRevB.82.134106} {\bibfield  {journal} {\bibinfo
  {journal} {Phys. Rev. B}\ }\textbf {\bibinfo {volume} {82}},\ \bibinfo
  {pages} {134106} (\bibinfo {year} {2010})}\BibitemShut {NoStop}%
\bibitem [{\citenamefont {Nishimatsu}\ \emph {et~al.}(2012)\citenamefont
  {Nishimatsu}, \citenamefont {Aoyagi}, \citenamefont {Kiguchi}, \citenamefont
  {J.~Konno}, \citenamefont {Kawazoe}, \citenamefont {Funakubo}, \citenamefont
  {Kumar},\ and\ \citenamefont {V.~Waghmare}}]{nishimatsu_molecular_2012}%
  \BibitemOpen
  \bibfield  {author} {\bibinfo {author} {\bibfnamefont {T.}~\bibnamefont
  {Nishimatsu}}, \bibinfo {author} {\bibfnamefont {K.}~\bibnamefont {Aoyagi}},
  \bibinfo {author} {\bibfnamefont {T.}~\bibnamefont {Kiguchi}}, \bibinfo
  {author} {\bibfnamefont {T.}~\bibnamefont {J.~Konno}}, \bibinfo {author}
  {\bibfnamefont {Y.}~\bibnamefont {Kawazoe}}, \bibinfo {author} {\bibfnamefont
  {H.}~\bibnamefont {Funakubo}}, \bibinfo {author} {\bibfnamefont
  {A.}~\bibnamefont {Kumar}}, \ and\ \bibinfo {author} {\bibfnamefont
  {U.}~\bibnamefont {V.~Waghmare}},\ }\href {\doibase 10.1143/JPSJ.81.124702}
  {\bibfield  {journal} {\bibinfo  {journal} {J. Phys. Soc. Jpn.}\ }\textbf
  {\bibinfo {volume} {81}},\ \bibinfo {pages} {124702} (\bibinfo {year}
  {2012})}\BibitemShut {NoStop}%
\bibitem [{\citenamefont {Grünebohm}\ \emph {et~al.}(2023)\citenamefont
  {Grünebohm}, \citenamefont {Teng},\ and\ \citenamefont
  {Marathe}}]{grunebohm_influence_2023}%
  \BibitemOpen
  \bibfield  {author} {\bibinfo {author} {\bibfnamefont {A.}~\bibnamefont
  {Grünebohm}}, \bibinfo {author} {\bibfnamefont {S.-H.}\ \bibnamefont
  {Teng}}, \ and\ \bibinfo {author} {\bibfnamefont {M.}~\bibnamefont
  {Marathe}},\ }\href {\doibase 10.1088/2515-7655/acd86f} {\bibfield  {journal}
  {\bibinfo  {journal} {JPhys Energy}\ }\textbf {\bibinfo {volume} {5}},\
  \bibinfo {pages} {034010} (\bibinfo {year} {2023})}\BibitemShut {NoStop}%
\bibitem [{\citenamefont {Walizer}\ \emph {et~al.}(2006)\citenamefont
  {Walizer}, \citenamefont {Lisenkov},\ and\ \citenamefont
  {Bellaiche}}]{walizer_finite-temperature_2006}%
  \BibitemOpen
  \bibfield  {author} {\bibinfo {author} {\bibfnamefont {L.}~\bibnamefont
  {Walizer}}, \bibinfo {author} {\bibfnamefont {S.}~\bibnamefont {Lisenkov}}, \
  and\ \bibinfo {author} {\bibfnamefont {L.}~\bibnamefont {Bellaiche}},\ }\href
  {\doibase 10.1103/PhysRevB.73.144105} {\bibfield  {journal} {\bibinfo
  {journal} {Phys. Rev. B}\ }\textbf {\bibinfo {volume} {73}},\ \bibinfo
  {pages} {144105} (\bibinfo {year} {2006})}\BibitemShut {NoStop}%
\bibitem [{\citenamefont {Nishimatsu}\ \emph {et~al.}(2008)\citenamefont
  {Nishimatsu}, \citenamefont {Waghmare}, \citenamefont {Kawazoe},\ and\
  \citenamefont {Vanderbilt}}]{Nishimatsu.2008}%
  \BibitemOpen
  \bibfield  {author} {\bibinfo {author} {\bibfnamefont {T.}~\bibnamefont
  {Nishimatsu}}, \bibinfo {author} {\bibfnamefont {U.~V.}\ \bibnamefont
  {Waghmare}}, \bibinfo {author} {\bibfnamefont {Y.}~\bibnamefont {Kawazoe}}, \
  and\ \bibinfo {author} {\bibfnamefont {D.}~\bibnamefont {Vanderbilt}},\
  }\href {\doibase 10.1103/PhysRevB.78.104104} {\bibfield  {journal} {\bibinfo
  {journal} {Phys. Rev. B}\ }\textbf {\bibinfo {volume} {78}},\ \bibinfo
  {pages} {104104} (\bibinfo {year} {2008})}\BibitemShut {NoStop}%
\bibitem [{\citenamefont {Bond}\ \emph {et~al.}(1999)\citenamefont {Bond},
  \citenamefont {Leimkuhler},\ and\ \citenamefont {Laird}}]{Bond.1999}%
  \BibitemOpen
  \bibfield  {author} {\bibinfo {author} {\bibfnamefont {S.~D.}\ \bibnamefont
  {Bond}}, \bibinfo {author} {\bibfnamefont {B.~J.}\ \bibnamefont
  {Leimkuhler}}, \ and\ \bibinfo {author} {\bibfnamefont {B.~B.}\ \bibnamefont
  {Laird}},\ }\href {\doibase 10.1006/jcph.1998.6171} {\bibfield  {journal}
  {\bibinfo  {journal} {J. Comput. Phys.}\ }\textbf {\bibinfo {volume} {151}},\
  \bibinfo {pages} {114} (\bibinfo {year} {1999})}\BibitemShut {NoStop}%
\bibitem [{\citenamefont {Vanderbilt}\ and\ \citenamefont
  {Cohen}(2001)}]{vanderbilt_monoclinic_2001}%
  \BibitemOpen
  \bibfield  {author} {\bibinfo {author} {\bibfnamefont {D.}~\bibnamefont
  {Vanderbilt}}\ and\ \bibinfo {author} {\bibfnamefont {M.~H.}\ \bibnamefont
  {Cohen}},\ }\href {\doibase 10.1103/PhysRevB.63.094108} {\bibfield  {journal}
  {\bibinfo  {journal} {Phys. Rev. B}\ }\textbf {\bibinfo {volume} {63}},\
  \bibinfo {pages} {094108} (\bibinfo {year} {2001})}\BibitemShut {NoStop}%
\bibitem [{\citenamefont {Nishimatsu}\ \emph {et~al.}(2013)\citenamefont
  {Nishimatsu}, \citenamefont {A.~Barr},\ and\ \citenamefont
  {P.~Beckman}}]{nishimatsu_direct_2013}%
  \BibitemOpen
  \bibfield  {author} {\bibinfo {author} {\bibfnamefont {T.}~\bibnamefont
  {Nishimatsu}}, \bibinfo {author} {\bibfnamefont {J.}~\bibnamefont {A.~Barr}},
  \ and\ \bibinfo {author} {\bibfnamefont {S.}~\bibnamefont {P.~Beckman}},\
  }\href {\doibase 10.7566/JPSJ.82.114605} {\bibfield  {journal} {\bibinfo
  {journal} {J. Phys. Soc. Jpn.}\ }\textbf {\bibinfo {volume} {82}},\ \bibinfo
  {pages} {114605} (\bibinfo {year} {2013})}\BibitemShut {NoStop}%
\bibitem [{Note1()}]{Note1}%
  \BibitemOpen
  \bibinfo {note} {We find $\Delta \protect \vec P$ for PbTiO$_3$ being
  ($+35$,$0$,$0$) in units of $\mu \protect \text {C/cm}^2$.}\BibitemShut
  {Stop}%
\bibitem [{Note2()}]{Note2}%
  \BibitemOpen
  \bibinfo {note} {At the second transition of BaTiO$_3$ Eqn.~\protect \eqref
  {eq:landau} predicts that $T_C^{\protect \mathrm {FC}}<T_C^{\protect \mathrm
  {ZFC}}$ within 22$^{\circ }$ away from $[100]$ on the $(0\protect \bar 11)$
  plane and within 16$^{\circ }$ away from $[100]$ on the $(001)$ plane, while
  MD predicts a smaller angle range (8$^{\circ }$ and 5$^{\circ }$). At the
  third transition, it predicts that $T_C^{\protect \mathrm {FC}}>T_C^{\protect
  \mathrm {ZFC}}$ within 19$^{\circ }$ away from $[111]$ on the $(\protect \bar
  110)$ plane and within 35$^{\circ }$ away from $[111]$ on the $(0\protect
  \bar 11)$ plane, while MD predicts a larger angle range (28$^{\circ }$ and
  51$^{\circ }$, respectively).}\BibitemShut {Stop}%
\end{thebibliography}%

\section{Appendix} 
\begin{figure}[h]
    \centering
    \includegraphics[width=0.48\textwidth, keepaspectratio]{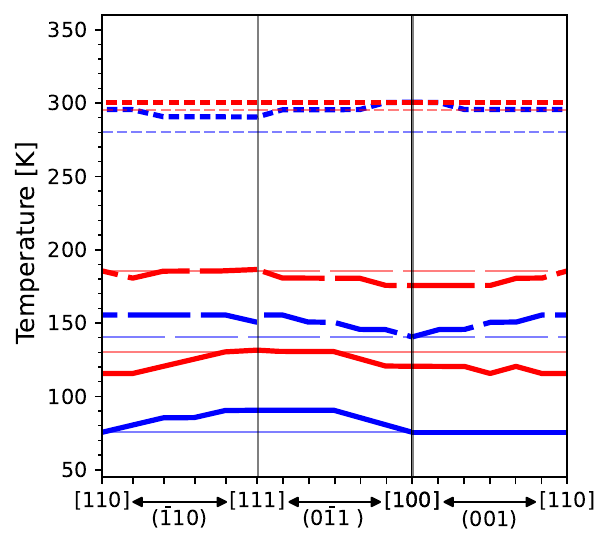}
    \caption{Temperature-electric field direction phase diagram of BaTiO$_3$ under fields of 10~kV/cm. Thick blue and red curves mark $T_C^{\mathrm{FC}}$ and $T_C^{\mathrm{FH}}$, 
and horizontal lines give $T_C^{\mathrm{ZFC}}$ and $T_C^{\mathrm{ZFH}}$ as reference. 
The change in $T_C$ and the reduction in thermal hysteresis \ch{show the same trends as for 100~kV/cm with smaller magnitude}. For example, at the third transition for field directions between $[100]$ and $[110]$, the change in $T_C^{\mathrm{FC}}$ is below the temperature resolution of 5~K. Note that the discontinuity of $T_C$, e.g.,\ at the second transition along field direction $[111]$, is an artifact of the 5~K temperature resolution. 
}
    \label{fig:phasediagram_BaTiO$_3$_10kVcm}
\end{figure}

\begin{figure}
    \centering
    \includegraphics[width=0.48\textwidth, keepaspectratio]{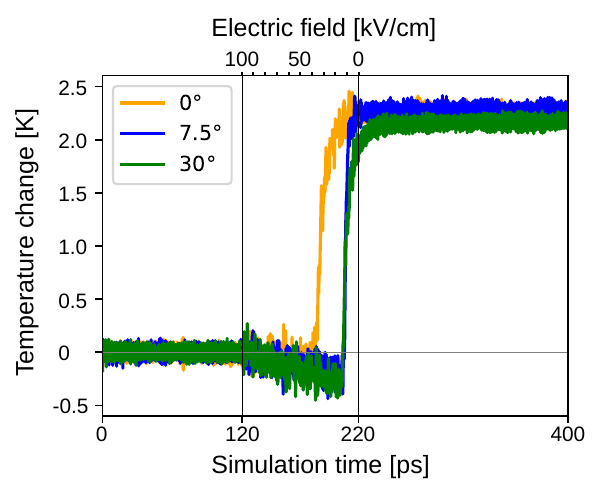}
    \caption{Time evolution of temperature changes relative to the initial temperature (i.e.,\ $T-T_{\text{in}}$) of about 70~K at the third transition 
  under field removal using a moving average of over 100 data points. Between 120 and 220~ps, the field is slowly ramped down from 100 to 0~kV/cm. Colors encode angles of the applied fields away from $[100]$ on the $(001)$ plane. During the field removal, the temperature, first, decreases due to the conventional ECE before the transitions (e.g.,\ 120 to 200~ps for 30$^{\circ}$) and then increases due to the inverse ECE at the transitions. 
   The larger the critical field strength for transition, the smaller the conventional contribution, the inverse ECE at the transition depends on the initial and final state.
     }
    \label{fig:partialcancel}
\end{figure}

\begin{figure*}
    \centering
    \includegraphics[width=1\textwidth, keepaspectratio]{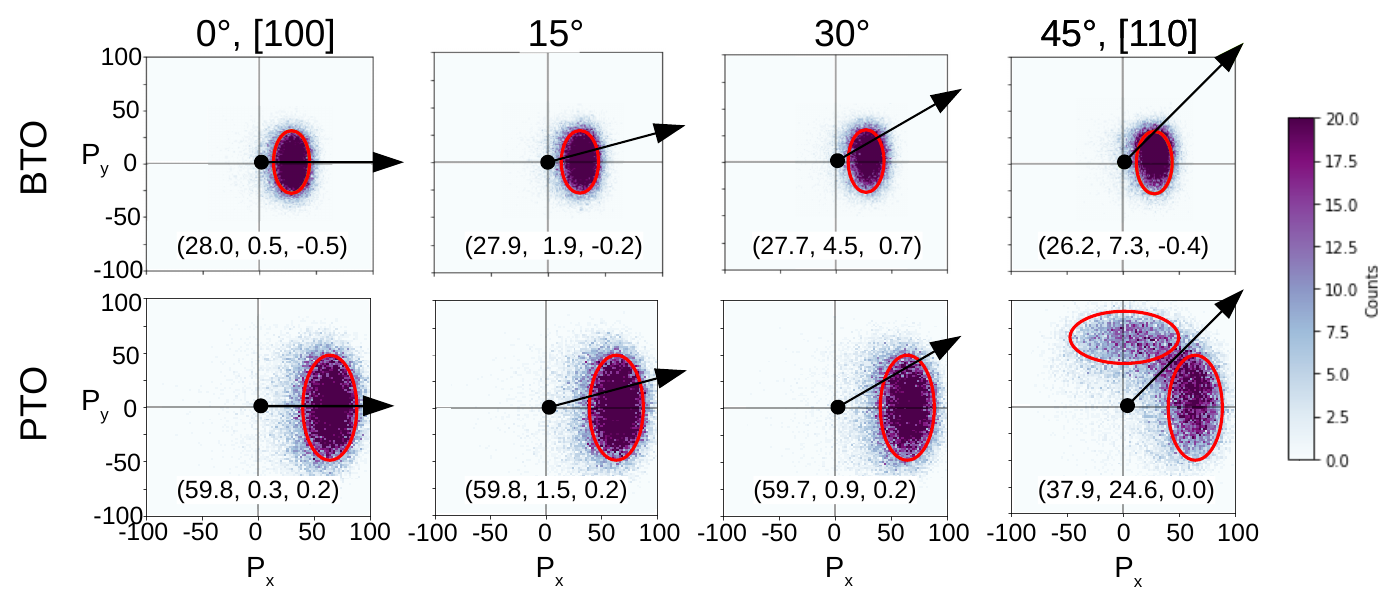}
    \caption{Polarization distribution of BaTiO$_3$ (at 300~K) and PbTiO$_3$ (at 640~K) below $T_C^{\mathrm{FC}}$ on the $P_x-P_y$ plane for field directions on the $(001)$ plane. The macroscopic polarizations (i.e.,\ the average over the whole system) are annotated in units of $\mu \text{C/cm}^2$. BaTiO$_3$ remains single-domain for all directions. For the $(100)$ direction, $P_x$ fluctuates between $-20$ and $60~\mu\text{C/cm}^2$, and $P_y$ and $P_z$ fluctuate between $\pm40 \mu \text{C/cm}^2$. When the field rotates away from $[100]$ to $[110]$, the field-induced polarization in $P_y$ increases. As the total polarization is the sum of the spontaneous polarization (along $[100]$) and the field-induced polarization, the polarization peak does not coincide exactly with the applied field direction. PbTiO$_3$ remains single-domain for fields deviating within 30$^{\circ}$ away from $[100]$ and the field-induced polarization is smaller compared to  BaTiO$_3$ due to a stronger polarization-strain coupling\cite{nishimatsu_molecular_2012}, and the width of the peaks is larger due to the higher temperature. For $[100]$ direction, $P_x$ fluctuates between $-$30 and 100 $\mu \text{C/cm}^2$, and $P_y$ and $P_z$ fluctuate between $\pm-75$ $\mu \text{C/cm}^2$. \ch{The time-averaged $P_z$ is zero.} For the field direction $[110]$, PbTiO$_3$ decomposes into a multidomain state whose macroscopic symmetry is M$_\text{C}$.   }
    \label{fig:Pdist}
\end{figure*}

In this appendix, we provide information on the electric field direction dependence under a weaker field (Fig.~\ref{fig:phasediagram_BaTiO$_3$_10kVcm}), the time evolution of temperature changes during field removal (Fig.~\ref{fig:partialcancel}), and the local dipole distribution for BaTiO$_3$ and PbTiO$_3$ (Fig.~\ref{fig:Pdist}).

\clearpage

\end{document}